\documentclass[sigconf]{acmart}
\usepackage{array}
\usepackage{booktabs}

\usepackage{graphicx}
\usepackage{textcomp}
\usepackage{xcolor}
\usepackage{xspace}
\usepackage{soul}
\usepackage{graphicx}
\usepackage{colortbl}
\usepackage[most]{tcolorbox}

\usepackage{listings}
\usepackage{verbatimbox}
\usepackage{threeparttable}
\usepackage{multirow} 
\usepackage{algorithmic}
\usepackage{graphicx}
\usepackage{multirow}
\usepackage{color,soul}
\usepackage{textcomp}
\usepackage{xcolor}
\usepackage{todonotes}
\usepackage{listings}
\usepackage{wrapfig}
\usepackage{tabularx}
\usepackage{url}
\usepackage{caption}
\usepackage{subcaption}
\usepackage{beramono}
\usepackage{tcolorbox}

\usepackage{xcolor}

\newcommand{\revised}[1]{{#1}}

\definecolor{buse_skyblue}{rgb}{0.337, 0.706, 0.914}

\definecolor{buse_orange}{rgb}{0.902, 0.624, 0.0}
\definecolor{buse_plum}{rgb}{0.7, 0.3, 0.5}

\definecolor{buse_green}{rgb}{0.4, 0.6, 0.4}

\definecolor{buse_mustard}{rgb}{0.8, 0.7, 0.2} 
\definecolor{buse_red}{rgb}{0.8392, 0.3921, 0.2666}
\definecolor{buse_purple}{rgb}{0.8, 0.475, 0.655}

\definecolor{buse_seafoam}{rgb}{0.5, 0.75, 0.6}
\definecolor{buse_blue}{rgb}{0.3333, 0.3764, 0.6627}

\definecolor{HAI}{HTML}{f4bbba}
\definecolor{HHAIshared}{HTML}{8dc1b8}
\definecolor{HHAIpersonal}{HTML}{a7c3e4}

\newtcbox{\CustomHHAIshared}{on line,
  arc=3pt, outer arc=3pt,
  colback=HHAIshared!50!white, colframe=HHAIshared!50!black,
  boxsep=.5pt, left=2pt, right=2pt, top=2pt, bottom=2pt,
  boxrule=0pt}

\newtcbox{\CustomHHAIpersonal}{on line,
  arc=3pt, outer arc=3pt,
  colback=HHAIpersonal!50!white, colframe=HHAIpersonal!50!black,
  boxsep=.5pt, left=2pt, right=2pt, top=2pt, bottom=2pt,
  boxrule=0pt}  

\newtcbox{\CustomHAI}{on line,
  arc=3pt, outer arc=3pt,
  colback=HAI!50!white, colframe=HAI!50!black,
  boxsep=0pt, left=2pt, right=2pt, top=2pt, bottom=2pt,
  boxrule=0pt}

\AtBeginDocument{%
  }

\setcopyright{acmlicensed}
\copyrightyear{2026}
\acmYear{2026}
\acmDOI{XXXXXXX.XXXXXXX}
\acmConference[CHI '26]{CHI Conference on Human Factors in Computing Systems}{April 13--17,
  2026}{Barcelona, Spain}

\begin{document}

\title[Human-Human-AI Triadic Programming]{Human-Human-AI \revised{Triadic} Programming: Uncovering the Role of \revised{AI Agent} and the Value of Human Partner in Collaborative Learning}


\author{Taufiq Daryanto}
\email{taufiqhd@vt.edu}
\affiliation{
  \institution{Virginia Tech}
  \city{Blacksburg}
  \state{VA}
  \country{USA}
}

\author{Xiaohan Ding}
\email{xiaohan@vt.edu}
\affiliation{
  \institution{Virginia Tech}
  \city{Blacksburg}
  \state{VA}
  \country{USA}
}

\author{Kaike Ping}
\email{pkk@vt.edu}
\affiliation{
  \institution{Virginia Tech}
  \city{Blacksburg}
  \state{VA}
  \country{USA}
}

\author{Lance T. Wilhelm}
\email{lancewilhelm@vt.edu}
\affiliation{
  \institution{Virginia Tech}
  \city{Blacksburg}
  \state{VA}
  \country{USA}
}

\author{Yan Chen}
\email{ych@vt.edu}
\affiliation{
  \institution{Virginia Tech}
  \city{Blacksburg}
  \state{VA}
  \country{USA}
}

\author{Chris Brown}
\email{dcbrown@vt.edu}
\affiliation{
  \institution{Virginia Tech}
  \city{Blacksburg}
  \state{VA}
  \country{USA}
}

\author{Eugenia H. Rho}
\email{eugenia@vt.edu}
\affiliation{
  \institution{Virginia Tech}
  \city{Blacksburg}
  \state{VA}
  \country{USA}
}

\renewcommand{\shortauthors}{Daryanto et al.}

\begin{abstract}
\revised{As AI assistance becomes embedded in programming practice, researchers have increasingly examined how these systems help learners generate code and work more efficiently. However, these studies often position AI as a replacement for human collaboration and overlook the social and learning-oriented aspects that emerge in collaborative programming.} Our work introduces \revised{human-human-AI (HHAI) triadic programming}, where an \revised{AI agent} serves as an additional collaborator rather than a substitute for a human partner. Through a within-subjects study with 20 participants, we show that triadic collaboration enhances collaborative learning and social presence compared to the dyadic human–AI (HAI) baseline. In the triadic HHAI conditions, participants relied significantly less on AI generated code in their work. This effect was strongest in the HHAI-shared condition, where participants had an increased sense of responsibility to understand AI suggestions before applying them. \revised{These findings demonstrate how triadic settings activate socially shared regulation of learning by making AI use visible and accountable to a human peer, suggesting that AI systems that augment rather than automate peer collaboration can better preserve the learning processes that collaborative programming relies on.}


\end{abstract}


\keywords{Do, Not, Us, This, Code, Put, the, Correct, Terms, for,
  Your, Paper}
\begin{teaserfigure}
  \centering
  \includegraphics[width=.9\textwidth]{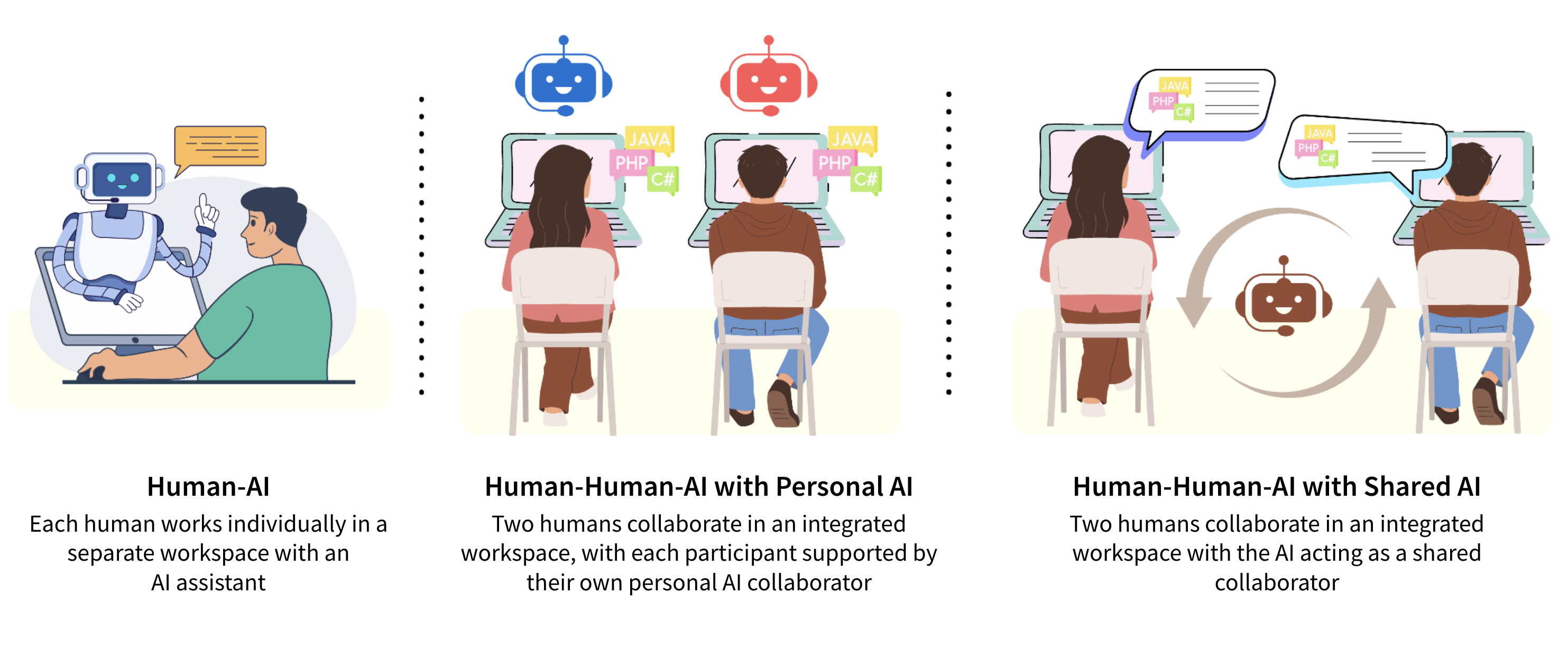}
  \caption{\textbf{Rethinking AI for Programming Learning.} Rather than replacing human partners, can AI augment a human–human pair programming while preserving the social and pedagogical benefits of collaboration? To answer that, we explore ``human–human–AI triadic programming'', where two humans work together with an \revised{AI agent}. Our study compares this approach to human–AI pairs and examines design considerations in this triadic interaction, including whether AI should act as a shared collaborator or personal support.}
  \label{fig:teaser-chi}
\end{teaserfigure}
\received{20 February 2007}
\received[revised]{12 March 2009}
\received[accepted]{5 June 2009}

\maketitle

\section{Introduction}
\revised{AI assistance has become increasingly prevalent in programming, with tools such as ChatGPT  \cite{openai2025chatgpt} and GitHub Copilot \cite{github_copilot} now common in everyday development practices \cite{weisz2025examining}. Prior work shows that these tools can accelerate productivity and improve task completion \cite{weisz2025examining}, and this trend has extended into educational settings where AI is frequently used for programming activities \cite{wang2023exploring}. Many studies adopt a human–AI pair programming setup in which a learner works with an AI model on the same task \cite{ma2023ai, fan2025impact}. These studies often position the AI as a replacement for a human partner \cite{ma2023ai}, and are frequently compared to collaborative programming practices such as human–human pair programming \cite{ma2023ai, fan2025impact}. Although substituting a peer with an AI can speed up progress and improve coding outcomes \cite{fan2025impact, ma2023ai}, this approach can overlook the social and collaborative dimensions that support learning in peer-based programming between two human partners \cite{fan2025impact}. This raises a central question for computing education about whether AI can augment human collaboration in ways that preserve the social and pedagogical benefits of peer learning.}

Existing studies largely examine the benefits of dyadic human–AI programming, with most work emphasizing speed and code quality as key evaluation metrics \cite{ma2023ai, fan2025impact}. \revised{However, these metrics tend to overlook the social and collaborative experiences in learning that shape how students learn \cite{nokes2015better}. In collaborative programming, learners work together by articulating their thought processes and sharing knowledge \cite{beck2013cooperative} as they jointly contribute to a shared task \cite{nosek1998case}}. \revised{Even in the era of AI, these collaborative skills remain valuable, particularly as modern work places increasingly adopt human-AI teaming practices \cite{hagemann2023human, nitsch2024human} where teams of humans and autonomous agents can work together toward shared goals \cite{berretta2023defining}. Students similarly rely on peers for shared reasoning, clarification, and support during programming activities \cite{wu2013enhancing}, yet they also recognize that AI can provide distinct forms of assistance that are difficult for peers to offer \cite{daryanto2025designing}. To better understand the combined strengths of both human and AI partners, we investigate a \textbf{human–human–AI triadic programming} approach where two human programmers collaborate with an AI agent as a third teammate.}

Moving from human-AI dyads to a triadic human–human–AI setting introduces distinct design considerations \cite{collins2024building}. First, should the AI act as a \textit{shared} collaborator for the human-human pair, or as \textit{personal} support for each human partner \cite{houde2025controlling, fahad2024role, zhu2024exploring, lyu2025will}? Second, beyond providing on-demand answers, what types of AI interventions are useful, for example, should the AI intervene proactively when the pair is stuck \cite{pu2025assistance, chen2025need, houde2025controlling}? \revised{Following recent work that defines proactive support as system-initiated assistance in the programming environment \cite{pu2025assistance}, we refer to a proactive AI agent as one that has the capability to initiate suggestions without always being explicitly prompted by the user.} Prior work shows that such proactive AI behaviors in programming settings can be perceived as helpful partners and improve collaboration experiences, but may also be disruptive \cite{pu2025assistance, chen2025need}. However, such work has focused exclusively on dyadic human--AI interactions. We know little about how a \revised{proactive AI agent} should participate in synchronous programming with two humans, where conversational flow, role coordination, and mutual accountability add additional layers of complexity \cite{hanks2011pair, zakaria2022two, bryant2008pair}. Clarifying when and how an AI should intervene, and how its positioning reshapes collaboration between two human partners, remains an open challenge. In this work, we address this gap through the design and evaluation of a \revised{human–human–AI triadic programming} system that allows us to compare shared versus personal AI positioning and proactive versus on-demand support.

To investigate this, we built an integrated workspace that connects two humans and an AI agent in real time, combining a collaborative code editor with a conversational interface that supports text and voice (Figure 2). The system is designed to support collaborative programming practice between the two humans, allowing them to actively engage in continuous dialogue to articulate their thought processes and coding. The AI agent complements this interaction and can assist on demand through hints and code analysis, as well as through lightweight proactive interventions (e.g., providing hints during idle periods), while avoiding premature solution reveal.

We then conducted a within-subjects study of 20 participants (10 pairs) to examine key design considerations for \revised{human-human-AI triadic programming}, such as the scope of AI interactions (shared collaborator versus personal support), the usefulness of different intervention types, and user perceptions towards \revised{human-human-AI triadic programming} in general. Specifically, we compared three conditions: (1) \textbf{Shared AI}, where the AI addresses the pair as a third collaborator; (2) \textbf{Personal AI}, where each human has an individual AI partner; and (3) \textbf{Human–AI}, a baseline condition where each human works alone with AI. 

Finally, we analyzed the usage of AI and participant perceptions across conditions.

In this study, we investigate:
\begin{itemize}
    \item \textbf{RQ1:} What are the benefits and drawbacks of \revised{human-human-AI triadic programming} compared to human-AI pair programming?
    \item \textbf{RQ2:} How does AI positioning (shared vs. personal) affect learning experiences in \revised{human-human-AI triadic programming}?
    \item \textbf{RQ3:} What types of AI support are useful for \revised{human-human-AI triadic programming}?
\end{itemize}

Our study shows that \revised{human-human-AI (HHAI) triadic programming} provides a stronger collaborative learning experience and greater social presence compared to the human–AI (HAI) condition. Participants noted that having both a human partner and an AI enabled multiple ways of learning, such as observing the interaction between their peer and the AI, or learning by explaining concepts to their partner. Interestingly, we also found that working with a peer (especially in the shared HHAI condition) increased participants’ sense of responsibility in how they engaged with the AI and corresponded with lower reliance on AI-generated code. Participants mentioned that knowing their partner was observing their interaction with the AI encouraged them to use the AI more responsibly.

We make the following contributions to HCI research on human–AI collaboration and learning:

\begin{enumerate}
    \item \textbf{Empirical evidence on triadic human–human–AI collaboration}. Through a within-subjects study with 20 computer science students, we provide the first systematic comparison of \revised{human–human–AI triadic programming} against human–AI baselines. Our findings show that adding a peer alongside AI restores collaborative learning and social presence, fosters accountable AI use, and reshapes how proactive AI suggestions are integrated.
    \item \textbf{A conceptual framing of augmentation versus automation in collaborative programming}. We show how replacing a peer with AI removes the interactional conditions that sustain dialogue, accountability, and explanation, whereas augmenting peer collaboration with AI preserves and reinforces these mechanisms. By grounding this distinction in empirical evidence from collaborative programming, we extend long-standing HCI discussion on automation versus augmentation with a concrete account of their divergent effects on collaborative learning.
    \item \textbf{Design principles for AI-supported collaborative programming}. We derive implications for future AI tools for collaborative programming: (a) making AI outputs visible to peers to scaffold accountability, (b) tuning AI proactivity to conversational flow, and (c) augmenting rather than displacing the pedagogical benefits of peer collaboration.
\end{enumerate}

\section{Related Work}


\subsection{AI for Programming Education}
The integration of AI into education has rapidly evolved in recent years \cite{gu2021predicting}, enabling more personalized learning experiences and supporting a wide range of applications \cite{ayeni2024ai}. These advancements have led to its adoption in various domains, including programming education \cite{phung2023generative}. For example, studies have examined how AI tools can assist students in coding tasks such as generating solutions, debugging, and learning programming concepts \cite{wang2023exploring, ghimire2024coding}. AI has also been applied to create programming education resources \cite{guo2023six}, such as generating coding exercises \cite{sarsa2022automatic, becker2023programming}. Additionally, researchers have developed interactive AI systems that tailor the learning process to specific computing skills \cite{jin2024teach, ma2024you, ma2024teach}. Furthermore, in higher education, several universities have begun integrating AI into programming courses by allowing the usage of AI coding assistants \cite{becker2023programming}.

\revised{Among these applications, AI integration has recently received increasing attention for its role in collaborative programming \cite{zhao2025generative, wang2025impact, tang2024vizgroup}. Collaborative programming refers to multiple programmers working jointly on the same coding task \cite{nosek1998case, wang2025impact}. One widely used form of this practice is pair programming \cite{ma2023ai, lyu2025will, fan2025impact}} where two programmers work together in a single workspace, in which one user assumes the \textit{driver} role of writing the code, while the other takes the \textit{navigator} role of reviewing and suggesting improvements \cite{hanks2011pair}. This practice has been shown to improve problem-solving skills \cite{williams2001support}, enhance code quality \cite{hulkko2005multiple}, and increase learning gains and satisfaction \cite{celepkolu2018importance}. \revised{In educational contexts, pair programming can also be done remotely \cite{bigman2021pearprogram}, where two students collaborate using a collaborative online IDE (Integrated Development Environment), while only one programmer writes code at a time, thus still reinforcing the driver-navigator framework. Studies also mention that remote pair programming can be similarly effective as in-person pair programming \cite{schenk2014distributed}.}

Recent work in this area has leveraged AI to support programming practice, primarily by substituting one human partner with an AI partner \cite{ma2023ai}. For example, studies have examined the use of tools such as GitHub Copilot or ChatGPT to assist with code writing, generate documentation, and produce comments \cite{lyu2025will}. Similarly, other research has explored AI for programming learning, where the AI provides hints or step-by-step guidance to support learners as they work through coding problems \cite{roest2024next}. More recent studies, such as \citet{pu2025assistance} and \citet{chen2025need}, have further advanced this approach by developing AI assistants capable of proactively providing help during the coding process.

While those AI-assisted programming approaches have shown benefits by increasing programming performance \cite{fan2025impact}, they often lack the social presence, mutual engagement, and peer support found in human–human collaborative programming \cite{fan2025impact}. Moreover, studies have shown that many students still prefer learning with a peer because they find it more enjoyable and experience a stronger human connection \cite{daryanto2025designing}. In collaborative programming in particular, as students collaborate through continuous dialogue, they develop their ability to articulate thought processes and problem-solving strategies \cite{beck2013cooperative}. Yet, despite these advantages, prior work has also noted the limitations of learning solely with another human partner, such as the lack of targeted, expert feedback \cite{daryanto2025designing}. Limiting collaboration to only human partners can therefore miss opportunities to leverage AI strengths to provide detailed support and feedback \cite{daryanto2025designing}.

To combine the benefits of both human and AI partners while addressing their respective limitations, we explore \revised{human–human–AI triadic programming}, where the AI acts as an additional collaborator rather than replacing a human partner. Although prior work has examined AI tools such as ChatGPT and GitHub Copilot for programming support \cite{lyu2025will}, little is known about how to effectively integrate AI into human–human–AI settings. Our study addresses this gap by investigating AI roles, types of beneficial AI interventions, and user perceptions in such collaborations.

\subsection{Multi-Human and AI Collaboration}
In recent years, human–AI collaboration has become an important research topic in HCI and CSCW \cite{fahnenstich2024trusting, chen2022human, wilhelm2025managers, zhang2023visar, desolda2025understanding}. Much of the existing work has focused on one-on-one interactions between humans and AI \cite{wang2024investigating, wilhelm2025managers, zhang2023visar, daryanto2025conversate}, where AI tools are designed to help individual users, such as writing assistance \cite{singh2024figura11y, zhang2023visar}, code completion \cite{desolda2025understanding, sun2025don}, learning \cite{wang2020human, daryanto2025conversate, wilhelm2025managers}, and decision support \cite{chen2022human, fahnenstich2024trusting}. These studies highlight the role of AI in supporting individuals and have driven the development of intelligent assistant applications.

In practice, however, many collaboration scenarios involve multiple people working together, such as group learning~\cite{cockrell2000context, javed2025exploring}, team brainstorming~\cite{muller2024group, gonzalez2024collaborative}, and decision-making~\cite{de2014team, chiang2024enhancing}. This has led researchers to explore multi-human–AI collaboration, examining the roles and functions of AI within group settings~\cite{javed2025exploring, he2024ai, chiang2024enhancing, houde2025controlling, zhang2025ladica}. For example, Houde et al. (2025) studied how conversational AI participates in small-group discussions \cite{houde2025controlling}. They found that AI could inspire and support groups but might also cause frustration when it dominated the conversation~\cite{houde2025controlling}. 


We observe that group collaboration with AI support often falls into two paradigms. The first is the \textbf{Personal AI} paradigm, where AI provides individualized feedback or suggestions to each group member, helping them contribute to the collective task~\cite{fahad2024role, zhu2024exploring}. For example, this occurs when individuals each use ChatGPT while still collaborating with one another~\cite{fahad2024role, zhu2024exploring}. The second is the \textbf{Shared AI Teammate} paradigm, where AI is designed to act as a group member that directly participates in collaboration alongside humans~\cite{houde2025controlling, muller2024group}. Houde et al.’s study illustrates this approach, showing that proactive AI participation can foster creativity and sustain conversational flow~\cite{houde2025controlling}.



These paradigms have been studied in various domains such as education~\cite{javed2025exploring}, design~\cite{he2024ai}, and group decision-making~\cite{he2024ai, chiang2024enhancing}. However, they remain underexplored in collaborative programming~\cite{lyu2025will, ma2023ai}. Drawing upon typical collaborative programming settings~\cite{bigman2021pearprogram, begel2008pair}, this form of collaboration is distinctive because it requires high synchronicity, with participants simultaneously engaging in coding and verbal communication while navigating different roles within the team~\cite{hanks2011pair, bryant2008pair, ma2023ai, fan2025impact}. This multimodal and multi-role nature sets collaborative programming apart from other collaboration contexts and raises open questions about how AI should participate. To address this gap, our work investigates how Personal AI and Shared AI paradigms shape multi-human–AI collaboration in the context of \revised{triadic programming}.





\section{System Design for \revised{Human–Human–AI Triadic Programming}}

\begin{figure*}[!h]
    \centering
    \includegraphics[width=\textwidth]{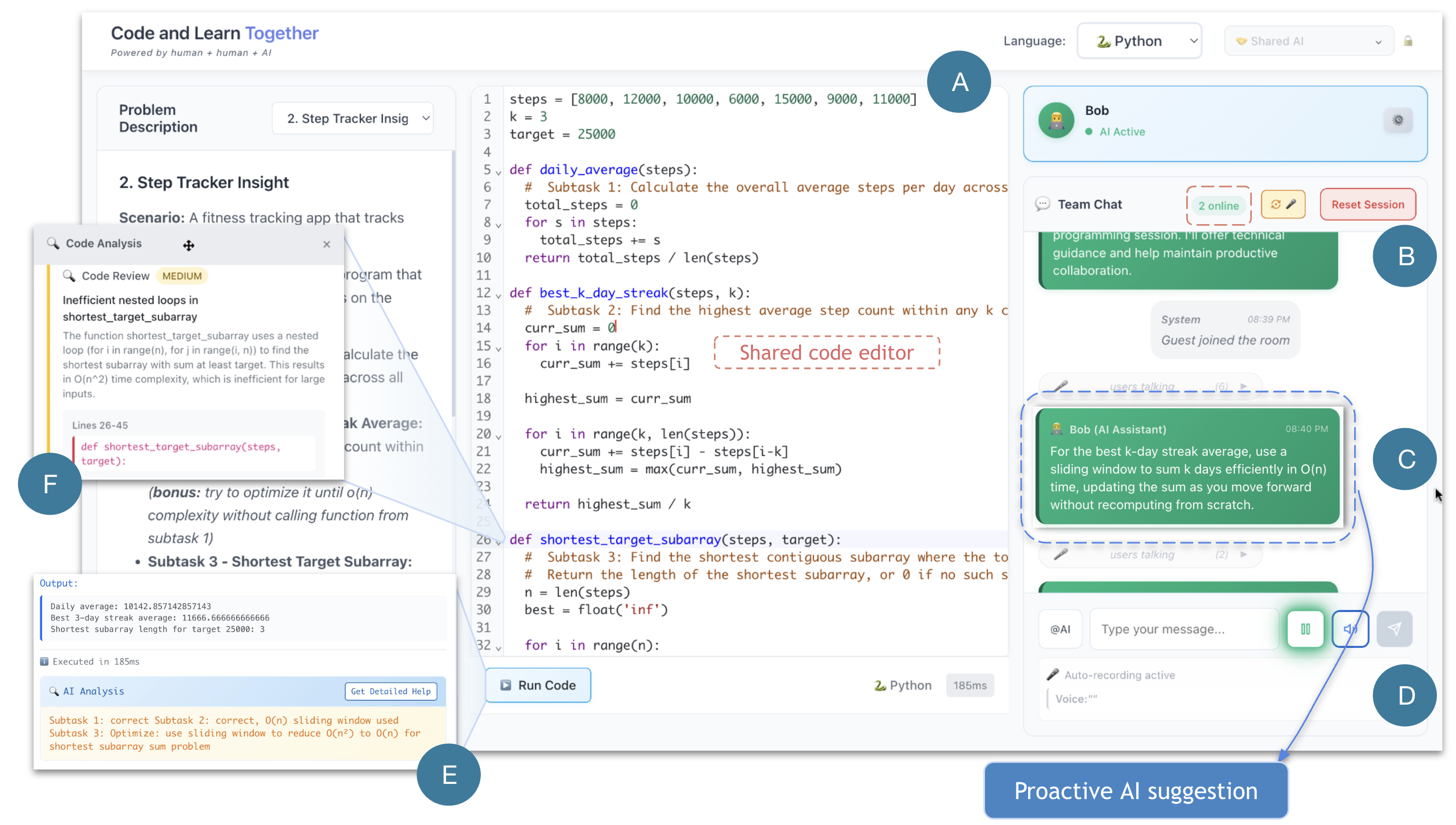}
    \caption{\textbf{Integrated Interface to Facilitate Human–Human–AI Triadic Programming.} \textbf{A) Collaborative Code Editor}: A live-shared editor where all participants can work together in real time.\textbf{ B) Conversational Interface}: Supports dialogue between the two humans and the AI. The AI can also be queried directly through spoken or typed questions \textbf{(Direct Request)}. \textbf{C) Proactive Intervention}: The AI agent can proactively intervene in the conversation with contextually relevant suggestions. \textbf{D) Live Transcription}: All spoken input is automatically transcribed using speech-to-text, allowing AI suggestions to be grounded in the ongoing conversation, the typed code, and the given problem. \revised{\textbf{E) Code Run Feedback}}: Each time the code is executed, the AI analyzes the output and provides debugging tips or improvement suggestions when necessary. \revised{\textbf{F) Code Block Analysis}}: By right-clicking a code block (e.g., a loop or function), users can request AI feedback that evaluates correctness and suggests improvements.}
    \label{fig:system_image}
\end{figure*}

We developed a system to support human–human–AI triadic programming.  The goal is to facilitate collaborative programming between two humans and an AI agent in an integrated environment where all participants can contribute simultaneously. The system is designed to enable collaborative learning between the two humans, enabling them to actively engage in continuous dialogue through which they articulate their thought processes and problem-solving strategies. The AI agent complements this interaction by proactively joining the discussion, providing guidance, offering hints, and suggesting potential solutions when needed.  

This system builds upon prior work on programming assistance \cite{lyu2025will, ma2023ai, roest2024next, pu2025assistance, chen2025need, zhao2025codinggenie} and AI-assisted learning \cite{daryanto2025designing, ma2024teach, ma2024you}, which emphasizes integration \cite{lyu2025will}, multimodality \cite{gupta2025multimodal}, and proactivity \cite{pu2025assistance, chen2025need, zhao2025codinggenie}. Our implementation enables two human programmers and an AI agent to collaborate in the same integrated workspace, supporting both human–human coordination and direct human–AI interaction to support collaboration and learning. The system supports multimodal interaction, allowing collaboration through voice, text, code typing, and interactive controls. The AI agent is designed to be proactive, meaning it can autonomously participate in the collaboration and provide assistance \cite{pu2025assistance, chen2025need}, while also remaining available on demand when explicitly queried. By using this system, we aim to explore \revised{human–human–AI triadic programming} for learning, enabling both humans to learn together with the support of a \revised{proactive AI agent}.

\revised{To clarify the boundary of our system’s capabilities, we designed our AI agent to be proactive but not fully agentic \cite{sapkota2025ai} because its interventions are based on predefined, task-specific rules, such as providing support after a period of user inactivity, rather than on autonomous decision making. Our design follows prior work on proactive AI agents that offer system-initiated assistance during idle moments in programming activity \cite{pu2025assistance, chen2025need}, as we describe in \S 3.2.1. Although the AI in our system is not fully agentic, prior work shows that users often perceived proactive AI agents as partners or collaborators rather than as simple tools \cite{pu2025assistance, houde2025controlling}.}

\subsection{Integrated Interface for \revised{Triadic Programming}}
The interface integrates a collaborative code editor with a conversational interface, and connects two humans with an AI agent simultaneously through WebSocket \cite{mdn_websockets}. This design supports three-way communication, allowing both humans and the AI to interact within the same environment. Interaction with the AI is multimodal: users can engage not only through text-based chat but also via voice, code editor, and interactive buttons. By keeping all interactions within a single workspace, we aim to reduce the need for context switching, addressing limitations noted in prior work where students had to move between the coding environment and external tools like ChatGPT during collaborative programming \cite{lyu2025will}.

\revised{Our tool is primarily designed to support remote collaboration. We focused on remote settings because they provide greater flexibility for participants to work together from different locations \cite{bolboacua2021practical}. Remote collaborative programming has become increasingly common in both educational contexts \cite{bigman2021pearprogram, hughes2020remote} and professional environments \cite{bolboacua2021practical}, offering benefits comparable to in-person settings \cite{schenk2014distributed}.}



\subsubsection{Collaborative Code Editor}
The code editor (Figure \ref{fig:system_image}. A) design was based on LeetCode \cite{leetcode}, a widely used platform for practicing programming. The interface displays the programming problem on the side and provides a code editor implemented using CodeMirror \cite{CodeMirror}. The editor supports multiple programming languages (C++, Java, and Python) and allows users to write and run code directly within the environment. We implemented real-time collaboration features that allow both users to edit code in the same online code editor, with support for viewing each other’s cursors and text highlights, similar to the Live Share feature in Visual Studio Code \cite{vscode_liveshare}. This synchronization was built using WebSocket \cite{mdn_websockets}, which enables two-way communication so that editing events such as keystrokes and cursor movements are shared instantly between participants with minimal delay.

\subsubsection{Conversational Interface}
We implemented a conversational interface into the workspace (Figure \ref{fig:system_image}. B), positioned on the right side of the coding environment. The interface supports both text and voice interaction, enabling communication between the two humans as well as with the AI agent. All spoken input is automatically transcribed using live speech-to-text transcription, allowing the AI to process dialogue in real time. Based on these transcripts, the AI can respond proactively or reactively according to several triggers, which we describe in the following subsection on AI interventions.

To support interaction, we also provided an option for users to enable AI voice output in addition to text \cite{daryanto2025designing}. Furthermore, every interaction with the AI is contextually grounded as the prompts include the current programming problem, the live state of the shared code editor, and the ongoing conversation. This design aims to provide relevant suggestions without requiring users to repeatedly explain their coding context \cite{lyu2025will}.

\subsection{AI Interventions}
We implemented the AI agent based on the GPT-4.1-mini large language model (All the prompts are provided in Appendix A). The AI agent is designed to participate in the collaboration through a set of intervention mechanisms informed by prior work on programming assistance \cite{pu2025assistance, chen2025need, mowar2025codea11y, zhao2025codinggenie} and AI-supported learning in programming \cite{roest2024next, daryanto2025designing}. These interventions are intended to provide support while maintaining opportunities for collaborative problem-solving among both users. We designed the AI agent to be able to intervene proactively in the conversation \cite{houde2025controlling}, or respond reactively when manually prompted or triggered by users \cite{roest2024next}. In addition, we prompted the LLM to support learning \cite{baidoo2023education} by offering guidance and hints without revealing full solutions too early \cite{roest2024next}, which aims to encourage peer discussion \cite{to2016making}.

\begin{enumerate}
    \item Proactive Intervention during Idle Period: The AI provides guidance or hints to re-engage users when both conversation and coding activity remain idle, without being explicitly prompted by users (Figure \ref{fig:system_image}. C). Productive collaboration relies on participants actively sharing their thoughts and engaging with one another \cite{de2024learning}, so extended silence is often unproductive \cite{edmondson2021reflections}. To address this, the AI intervenes by offering guidance that can help learners overcome moments of being stuck and continue making progress \cite{haindl2024students}. This design is informed by prior work on proactive programming assistance, which demonstrates the benefits of offering support during idle periods \cite{chen2025need, pu2025assistance}. Following prior implementations \cite{chen2025need}, the AI triggers interventions after 5 seconds of silence or inactivity (i.e., neither typing nor talking), but restricts them to once every 20 seconds since the last AI message to reduce disruptiveness.

    \item Direct Request: Users can directly engage the AI by addressing it explicitly (e.g., saying “Hey Bob …” (Bob is the name for the AI agent) or typing “@AI …”) to request clarification, hints, or additional explanations. They can also ask follow-up questions based on prior suggestions. To support the learning process, the AI is prompted not to reveal full solutions, but instead to provide just enough hints or guidance to help users make progress \cite{roest2024next}.

   \item Task Scaffolding: We implemented a code editor feature that generates inline suggestions from user comments, similar to interactions in existing AI coding assistants (e.g., GitHub Copilot \cite{github_copilot}). However, when users write inline instructions as comments in the code editor (e.g., “\#\# implement …”), the AI generates task scaffolding or step-by-step todos instead of directly providing the full code \cite{lin2021pdl}. This approach is designed to support the learning process by helping users get started without immediately revealing the solution. If users remain stuck, they can right-click on a todo line to reveal the corresponding code implementation for that specific line.
   
  \item Code Run Feedback: Each time the code is executed, the AI analyzes the output and provides suggestions or debugging tips when necessary (Figure \ref{fig:system_image}. E) \cite{pu2025assistance}. If users are still uncertain about the feedback, they can click “Get Detailed Help” to bring the AI’s explanation into the conversation for more detailed guidance.
  
    \item Code Block Analysis: By right-clicking a code block (e.g., a loop or a function), users can request an AI analysis that evaluates whether the implementation is correct and suggests possible improvements, such as enhancing code efficiency (Figure \ref{fig:system_image}. F). One potential use case of this feature is when a user finishes writing a block of code and wants to check its quality. Another potential use case is when the navigator (the other user who is not actively coding) wants to verify the implementation while reviewing the driver’s code, using the AI as additional support \cite{bryant2008pair}.

\end{enumerate}



\subsection{Human-Human-AI Interaction: Shared Versus Personal}
Drawing on prior work that examined different ways of integrating AI into collaborative settings \cite{houde2025controlling, he2024ai, lyu2025will, fahad2024role, zhu2024exploring}, we aim to explore how an AI agent can be positioned in human–human–AI triadic programming: either as a shared collaborator supporting both participants collectively \cite{houde2025controlling, zhang2025ladica, he2024ai}, or as a personal collaborator assisting each participant individually \cite{fahad2024role, zhu2024exploring}.

\subsubsection{\CustomHHAIshared{\textbf{Shared AI}}}
In this mode, the AI acts as a third collaborator, interacting with both humans simultaneously. The AI is positioned as a shared collaborator that can proactively engage with the group. All interventions are directed to the pair as a whole, enabling the AI to participate directly in the joint dialogue between the two human collaborators. This design builds on a growing interest in non-dyadic human–AI interaction \cite{collins2024building} and recent studies showing how proactive AI can support group discussions by contributing ideas and facilitating conversation \cite{houde2025controlling}. It also parallels findings in collaborative programming where third-party interventions, such as those from an instructor, have been shown to assist the learning process \cite{zakaria2022two}.

\subsubsection{\CustomHHAIpersonal{\textbf{Personal AI}}}
In this mode, each human is paired with their own AI collaborator. This design reflects common practices where individuals use AI tools (e.g., ChatGPT) while still collaborating with others \cite{fahad2024role, zhu2024exploring}. It also aligns with current programming conditions in which collaborators may each rely on tools such as ChatGPT or GitHub Copilot \cite{lyu2025will}. In our implementation, the AI can still provide proactive interventions, but it interacts only with its paired user rather than addressing the group. Unlike in the shared mode, the AI does not use voice output in this configuration because its interventions are directed only to each individual. Instead, the AI provides text-based support, making its assistance more individualized while the two humans continue to collaborate with each other.

\section{\revised{Empirical Study of Human-Human-AI Triadic Programming}}
\subsection{Participants}
We recruited 20 computer science students (10 pairs) from the first author’s university. \revised{Participants registered together as pairs who already knew each other. We acknowledge that this choice limits the generalizability of our findings, as working with unfamiliar partners may lead to different experiences, which we leave for future work to explore.}

All participants had experience using AI tools for coding (e.g., ChatGPT \cite{openai2025chatgpt}). The majority (18/20) had prior experience with collaborative programming, either through class projects or personal projects. Most of our participants (16/20), who were third-year undergraduates or above, had taken a data structures and algorithms course, while the remaining four participants (second-year students) had taken an intermediate data structures course. Participants were recruited through the university mailing list and class announcements. All participants were compensated \$25 for a 90-minute user study. Detailed demographic information is provided in Table~\ref{tab:participant_demographics}.

\begin{table*}[h]
\centering
\caption{Participant Demographics (Pairs)}
\label{tab:participant_demographics}
\begin{tabular}{|c|c|c|c!{\vrule width 1.5pt}c|c|c|c|}
\hline
\textbf{ID} & \textbf{Gender} & \textbf{Ethnicity} & \textbf{Education} & 
\textbf{ID} & \textbf{Gender} & \textbf{Ethnicity} & \textbf{Education} \\
\hline
P1  & Male   & Asian  & 3rd Year Undergrad & P2  & Male   & Asian  & 3rd Year Undergrad \\
\hline
P3  & Male   & Asian  & Master                 & P4  & Male   & White    & Master \\
\hline
P5  & Male   & Asian  & Master                 & P6  & Male   & Asian  & 1st Year PhD \\
\hline
P7  & Male   & Asian  & 2nd Year Undergrad               & P8  & Male   & Asian  & 2nd Year Undergrad \\
\hline
P9  & Male   & White  & 3rd Year Undergrad               & P10 & Male   & White  & 3rd Year Undergrad \\
\hline
P11 & Male   & White  & 2nd Year Undergrad               & P12 & Male   & MENA   & 2nd Year Undergrad\\
\hline
P13 & Male   & MENA   & Master                 & P14 & Female & White  & Master \\
\hline
P15 & Male   & Asian  & Master                 & P16 & Male   & Asian  & Master \\
\hline
P17 & Female & Asian  & 3rd Year Undergrad               & P18 & Female & White  & 3rd Year Undergrad\\
\hline
P19 & Male   & Black  & 3rd Year Undergrad               & P20 & Female & Black  & 3rd Year Undergrad\\
\hline
\end{tabular}
\end{table*}

\subsection{Study Design}
To examine the role of \revised{AI agent} and the value of a human partner in \revised{human–human–AI triadic programming}, we conducted a within-subject study in which participants engaged with our system across three different conditions (Figure \ref{fig:teaser-chi}). The order of conditions was counterbalanced to mitigate ordering and learning effects. All study sessions were conducted remotely using Zoom, as collaborative programming can also be done in a remote setting \cite{hughes2020remote, dominic2020remote}. \revised{In our study, although our tool allows for simultaneous code editing, we instructed participants that only one person should edit the code at a time, which grounds our work in the driver–navigator model of pair programming \cite{bigman2021pearprogram}. However, participants could switch roles at any point by coordinating verbally, since both participants had access to the same editor. We draw this practice from prior literature on remote pair programming, which suggests the benefit of maintaining the driver–navigator model \cite{adeliyi2021investigating}.}


The three study conditions were as follows:
\begin{enumerate}
\item \CustomHHAIshared{\textbf{Human–Human–AI with Shared AI}}: Two humans collaborate in real time in an integrated workspace with the AI acting as a shared collaborator (\S 3.3.1). In this mode, all AI interventions are directed to the pair collectively, positioning the AI as a third collaborator in the collaboration.  


\item \CustomHHAIpersonal{\textbf{Human–Human–AI with Personal AI}}: Two humans collaborate in real time in an integrated workspace, with each participant supported by their own personal AI collaborator (\S 3.3.2). Here, the AI provides individual support to each human, enabling personalized guidance while still working within the shared environment. Comparing the first and second conditions enables us to understand how to position AI in \revised{human-human-AI triadic programming}, either as a personal collaborator or a shared collaborator.

\item \CustomHAI{\textbf{Human–AI}}: Each human works individually in a separate room with an AI assistant, using the same implementation described in \S 3.3.1 but without a human partner. This ablated condition allows us to compare collaboration with and without a human partner, helping uncover the value and role of a human partner in the process, as well as whether users perceive the AI differently in human–AI versus human–human–AI collaboration.
\end{enumerate}

\revised{Our study focuses on conditions that involve AI because contemporary programming practices increasingly rely on AI-assistance. As a result, our baseline is a human–AI rather than a human–human collaboration setup. Although human–human collaboration in pair programming, has long been an established practice in computing education \cite{salleh2010empirical}, recent research has significantly shifted toward AI-assisted programming \cite{ma2023ai, lyu2025will, fan2025impact}. This shift reflects the extent to which AI tools have become integrated into both educational \cite{liu2024teaching} and professional programming contexts \cite{kumar2025sharp}, making their use inseparable from modern learning and development practices in programming contexts  \cite{liu2024teaching, wang2023exploring}.}


\revised{By exploring three AI-supported collaboration conditions, we aim to examine how people collaborate with AI under different configurations and understand how these configurations shape programming practice. Moreover, by comparing the human–human–AI and human–AI conditions, we seek to uncover the added value of having a human partner alongside the AI, highlighting how AI can complement human collaboration in the learning contexts. This can potentially reinforce the importance of maintaining human involvement rather than positioning AI as a replacement for human partners. }


\subsection{Programming Tasks}
The programming tasks were adapted from LeetCode problems \cite{leetcode}, as their scope and style are well-suited for studying programming practice in line with prior work \cite{daryanto2025designing, pu2025assistance}. Each task was modified to include multiple subproblems that participants could work on. We designed the tasks to support multiple solution approaches with varying levels of optimization, enabling participants to explore different strategies and reflect on trade-offs \cite{daryanto2025designing}. This flexibility encouraged exploration and discussion among both humans and the AI. We selected tasks that emphasized algorithmic thinking rather than heavy syntax, since developing strong problem-solving skills is critical for students’ reasoning abilities and highly relevant for career preparation \cite{daryanto2025designing}. To ensure fairness across conditions, all tasks were constructed to be of comparable difficulty.  Additionally, the coding tasks are distributed equally across all 3 conditions to avoid confounding task difficulty. All tasks are provided in Appendix B.

\subsection{Procedure}

The study was conducted online via Zoom, where two participants joined the session from different locations. We began by asking participants to complete a consent form, as the study had been approved by the Institutional Review Board (IRB) at the first author’s institution. Next, we introduced the tool and provided a live demonstration. Participants were then instructed to work on three conditions (\CustomHHAIshared{HHAI Shared}, \CustomHHAIpersonal{HHAI Personal}, and \CustomHAI{HAI} as shown in Figure~\ref{fig:teaser-chi}). The order of conditions was counterbalanced across participants. Each condition involved a separate problem. Before each task, participants received an explanation of the tool and the condition. They were given 17 minutes to complete each task. After finishing a task, participants completed a questionnaire. At the end of the study, we conducted a semi-structured interview to gather insights into their experiences and perceptions of each condition. In total, the user study lasted 90 minutes. Each participant was compensated with a $\$25$ Amazon gift card for the study.

\begin{figure*}[!h]
    \centering
    \includegraphics[width=0.7\textwidth]{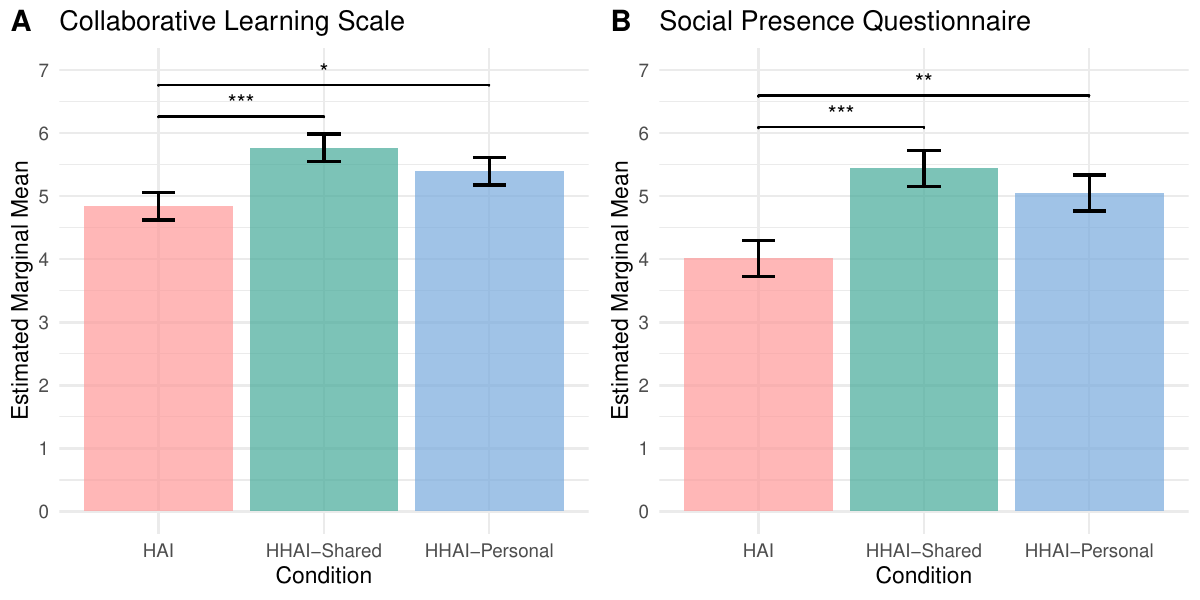}
    \caption{Collaborative Learning Scale (CLS) and Social Presence Questionnaire (SPQ): Both HHAI conditions showed significant differences compared to the HAI condition.}
    \label{fig:learning_presence}
\end{figure*}

\subsection{Data Collection and Analysis}

\textbf{Post-task questionnaire:} After finishing each task, participants completed a questionnaire. This included the Collaborative Learning Scale \cite{so2008student, fan2025impact}, the Social Presence Questionnaire \cite{kreijns2011measuring, fan2025impact}, and several individual questions about their perceptions of AI in the collaboration. All questions are provided in Appendix C. 

\textbf{Qualitative interview:} All study sessions were recorded on Zoom and automatically transcribed. At the end of the study, we conducted semi-structured interviews to gather participant insights. We then applied thematic analysis \cite{braun2021thematic}. The first author conducted inductive coding to generate codes grounded in the data \cite{braun2023toward}. \revised{Then, these codes were discussed with another author to refine the coding, and we constructed broader themes and connected them to the quantitative findings \cite{dixon2005synthesising}. We used the coding mainly to identify representative quotes that illustrate and contextualize the quantitative findings and to help inform the design implications by interpreting how participants experienced the AI during collaboration.}

\textbf{AI usage:} We captured all interaction data in our database, including conversation logs, code typed line by line, and AI interventions. We also recorded the sessions on Zoom while both participants shared their screens. Because Zoom only records the screen of the most recent participant to share, we additionally recorded the first participant’s screen using a screen recording app on our laptop (This was possible by toggling between shared screens and manually recording the desired view). To analyze AI usage, two researchers manually annotated each AI intervention. For each AI suggestion, we marked whether it was used or not. We defined “AI suggestion is used” as cases where participants (i) incorporated it into code (e.g., syntax fixes, code changes based on hints, or copy-pasted AI-generated code), (ii) discussed it with their partner, or (iii) asked follow-up questions to the AI. Annotations were made based on the interaction data in our database and aligned with Zoom-recorded events.

\textbf{AI-generated code usage:} In addition to tracking AI interventions, we also counted how many lines of code were fully generated by the AI versus written by participants. Syntax fixes were excluded since they involved partial human input. We counted a line as AI-generated if participants directly copy-pasted or manually wrote code from the AI suggestion into their solution fully.

\revised{\textbf{Number of sub-tasks completed:}}
\revised{To measure participants' programming performance, we followed prior approaches \cite{chen2025need} by counting how many sub-tasks they completed. Each problem contained three sub-tasks. Sub-task 1 had a single target solution, while sub-tasks 2 and 3 could be solved either with a brute-force solution or with an optimized solution.}

\subsubsection{Statistical Analysis}
All statistical analyses were conducted in R (version 4.5.1). Scores for the Collaborative Learning Scale (CLS) \cite{so2008student, fan2025impact} and the Social Presence Questionnaire (SPQ) \cite{kreijns2011measuring, fan2025impact} were calculated by averaging across items within each scale to yield a final composite score. For continuous outcome variables from the survey measures, we used linear mixed-effects models~\cite{oberg2007linear}. \revised{Mixed-effects models accounted for the repeated-measures structure of the study, where each participant completed all three conditions and participants worked in pairs. The models included fixed effects for condition and random intercepts for participant ID and group ID to capture individual- and pair-level variability.} The order of conditions was included as a fixed effect to account for potential ordering effects. For proportional outcome variables,  such as the proportion of AI-generated code and the proportion of AI suggestions used, we used the same models with a binomial error distribution. These models incorporated the same random and fixed effects structure described above. For individual survey items with 7-point Likert scale responses, cumulative link mixed models \cite{christensen2018cumulative} were used to appropriately account for the ordinal nature of the response variables. Post-hoc pairwise comparisons between experimental conditions were conducted using estimated marginal means \cite{searle1980population}. Across all analyses, the threshold for statistical significance was set at $\alpha = 0.05$



\section{Findings}

\subsection{HHAI Restores Collaborative Learning and Social Presence}

Collaborating with a human partner restored the collaborative learning and social presence that participants felt were missing in human--AI dyads, with the strongest gains in the \CustomHHAIshared{Shared AI} condition. As shown in Figure~\ref{fig:learning_presence}, both collaborative learning and social presence ratings were higher in HHAI than in HAI, with Shared AI producing the largest effects.  

\subsubsection{Supporting Collaborative Learning} Participants emphasized that working with another human provided opportunities to learn in ways absent in HAI. P3 described how simply \emph{``observing what my partner is doing''} helped them pick up new techniques. Others highlighted the value of explanation; P18 reflected that \emph{``when I talked to my partner... she would catch my mistakes that I didn’t even think twice about... just talking out loud [helps].''} Such accounts contrast with HAI sessions, where participants often stayed silent and engaged mainly with the AI’s outputs. 

Our analysis of the CLS questionnaire results revealed significant main effects of experimental condition on collaborative learning ratings ($F(2,37) = 7.31$, $p < .01$). Compared to the human--AI baseline (HAI; $M = 4.63$), participants reported significantly higher collaborative learning in both the HHAI--Shared condition ($\beta = 0.93$, $SE = 0.24$, $t(37) = 3.81$, $p < .001$) and in the HHAI--Personal condition ($\beta = 0.55$, $SE = 0.24$, $t(37) = 2.28$, $p = .028$) with no significance in the experimental order ($p = .381$). Participants noted that the benefit of both HHAI conditions came less from efficiency and more from being pushed to explain and refine their thinking in dialogue with a peer.

\subsubsection{Strengthening Social Presence} Participants consistently emphasized that collaborating with a peer felt more natural and socially engaging than interacting with AI alone. One participant explained:  

\begin{quote}
    ``I would say you can explain your thoughts to your partner in a more easier way. I think he can, like, try to understand what I’m trying to do, but conversing that to AI is harder for me. So, like, for me, I would just ask AI to, again, just solve it. So, sharing my thoughts loudly is easier to share with your partner, rather than AI.'' -- P5
\end{quote}

Other participants echoed this sentiment, describing AI-only work as \emph{``awkward''} (P7) or \emph{``one-sided''} (P12) compared to the back-and-forth with a peer. Social presence ratings showed a similar pattern of differences across conditions ($F(2,37) = 9.45$, $p < .001$). Relative to HAI ($M = 4.13$), participants in both HHAI conditions reported significantly higher social presence (HHAI--Shared: $\beta = 1.43$, $SE = 0.37$, $t(37) = 3.90$, $p < .001$; HHAI--Personal: $\beta = 1.04$, $SE = 0.37$, $t(37) = 2.84$, $p = .007$) with no significance in the experimental order ($p = .764$). Post hoc comparisons (Figure \ref{fig:learning_presence}.B) further indicated that HHAI–Shared did not differ from HHAI–Personal ($p > .05$). 

Overall, participants contrasted HHAI with HAI to emphasize how a human peer restored the pedagogical and social value of collaboration. Quantitative results confirmed that both HHAI conditions improved collaborative learning and social presence, and that Shared AI magnified these effects.


\subsection{HHAI Encourages More Accountable Use of AI}

Working with a peer, especially in the \CustomHHAIshared{Shared AI} condition, increased participants’ sense of responsibility for how they engaged with AI and corresponded with lower reliance on AI-generated code. As shown in Figure~\ref{fig:responsibility_reliance}, responsibility ratings were highest in Shared AI, and the proportion of AI-generated code was lowest in HHAI compared to \CustomHAI{HAI}.

\subsubsection{Heightened Responsibility to Understand Before Applying AI Suggestions} 

Participants explained that the presence of a human partner made them more careful about how they applied AI output. For instance, P6 explained that simply knowing a partner could see her prompts made her feel more accountable and careful in how she engaged with the AI:  
\begin{quote}
``If I see that someone is watching my AI prompts, I would feel that I need to be more mature and try to understand the code first, rather than just copy-pasting.'' -- P6
\end{quote}
Similarly, P13 also emphasized how this visibility created a sense of social pressure that encouraged them to be more deliberate about how they used AI:  
\begin{quote}
``There’s a little bit of pressure… I don’t want to be looked bad or looked down upon by being reliant on AI when coding with another human.'' -- P13
\end{quote}

\begin{figure*}[]
    \centering
    \includegraphics[width=0.8\textwidth]{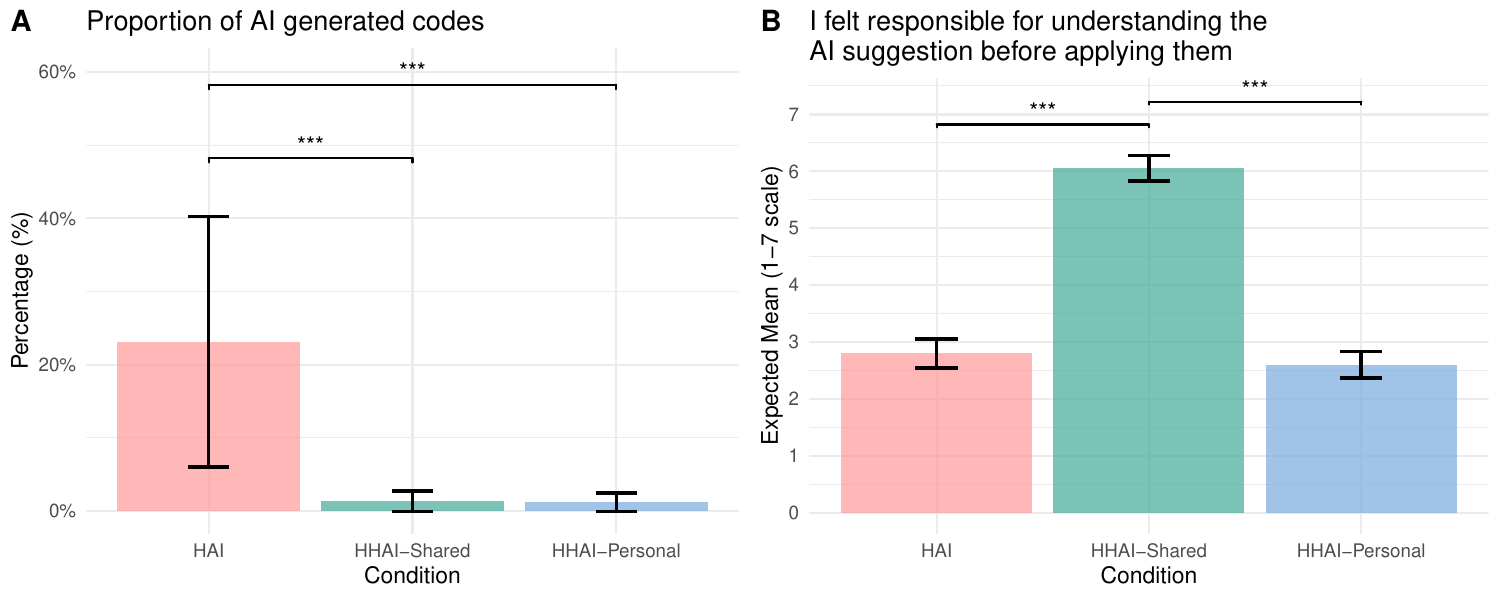}
    \caption{A) Proportion of AI-generated code; B) Perceived responsibility for understanding AI suggestions before applying}
    \label{fig:responsibility_reliance}
\end{figure*}

Our questionnaire results also reflected this pattern. Analysis using cumulative link mixed models revealed that the influence of a human partner on participants' sense of responsibility was most pronounced in the \CustomHHAIshared{HHAI--Shared} condition (Wald $\chi^{2} = 24.4$, $p < .001$). As shown in Figure~\ref{fig:responsibility_reliance}.B, HHAI--Shared produced the highest responsibility ratings ($M = 6.05$, 95\% CI [5.61, 6.49]) compared to HAI ($M = 2.80$, 95\% CI [2.30, 3.30]) or HHAI--Personal ($M = 2.60$, 95\% CI [2.15, 3.06]). Post-hoc comparisons indicated that HHAI--Shared generated significantly higher responsibility ratings than both HAI ($p < .001$) and HHAI--Personal ($p < .001$). However, HAI and HHAI--Personal did not differ significantly ($p = .819$). The analysis also revealed a significant negative order effect, with participants reporting a decreasing sense of responsibility in later tasks ($\beta = -0.95$, $z = -2.88$, $p = .004$).


\subsubsection{Decreased Reliance on AI-Generated Code} The same sense of accountability was evident in how often participants prompted the AI to generate code and how much of that output they incorporated into their final code submission. In \CustomHAI{HAI}, a larger share of the submitted code came from the AI. Our analysis using a generalized linear mixed model with binomial distribution showed a strong effect of condition (Wald $\chi^{2} = 48.4$, $p < .001$). As shown in Figure~\ref{fig:responsibility_reliance}.A, the HAI condition showed the highest proportion of AI-generated code (23.1\%, 95\% CI [4.4\%, 66.5\%]), while both HHAI conditions demonstrated substantially lower reliance on AI-generated code: HHAI--Shared (1.4\%, 95\% CI [0.2\%, 9.5\%]) and HHAI--Personal (1.2\%, 95\% CI [0.2\%, 8.9\%]). Experimental order showed a significant effect ($\beta = 0.63$, $z = 2.44$, $p = .015$), indicating increased AI code usage in later sessions.


Many participants connected this reduction in AI-generated code to the social dynamics of working with a human peer. However, this reduction was not universal. In one pair (P15 + P16), $56\%$ of their final code still came from the AI, despite being in the HHAI condition. P15 reflected on this reliance:  
\begin{quote}
``Because we have been so habituated to use AI rather than talking to a human, it just came out of habit.'' -- P15
\end{quote}
Their transcript also showed limited dialogue between the two partners, suggesting that weak human-to-human communication in collaborative programming undermined the accountability effect. 


Overall, participants described feeling more responsible for understanding AI suggestions when collaborating with a partner, and this was reflected in both higher responsibility ratings and lower proportions of AI-generated code. \CustomHHAIshared{Shared AI} in particular heightened accountability by making prompts and responses visible to both collaborators.

\subsection{Shared AI Supports Flow, While Personal AI Disrupts It}

\begin{figure*}[]
    \centering
    \includegraphics[width=0.8\textwidth]{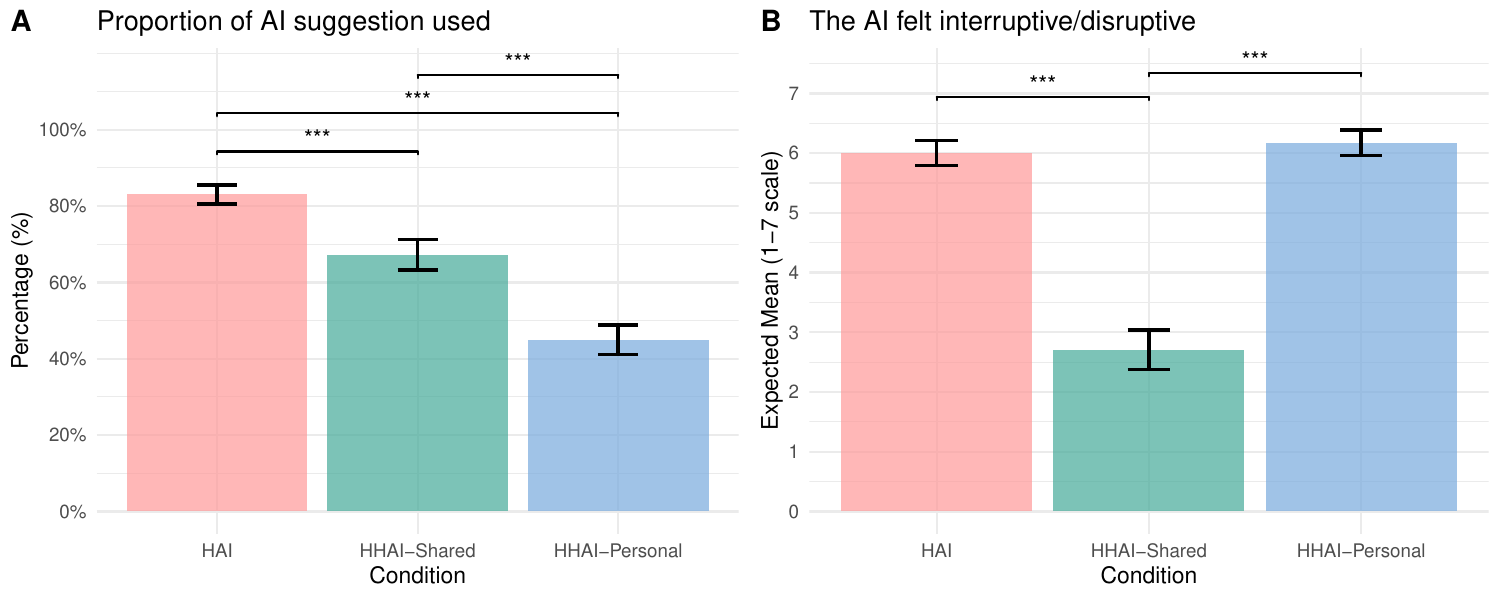}
    \caption{A) Proportion of AI suggestions being used (i.e., used in code, discussed, or followed up) based on annotated events; B) Questionnaire results on participants’ experiences about AI proactivity}
    \label{fig:used_suggestion_and_interruption}
\end{figure*}

Shared AI interventions were experienced as the least disruptive and best aligned with group flow, while Personal AI fragmented collaboration and proactive suggestions were often ignored. As shown in Figure~\ref{fig:used_suggestion_and_interruption} B, Shared AI received significantly lower disruptiveness ratings, and participants were far more likely to adopt directly requested suggestions than proactive ones.  

\subsubsection{Sense of AI Disruptiveness} A cumulative link mixed model showed that participant ratings of AI disruptiveness differed substantially across conditions (Wald $\chi^{2} = 23.5$, $p < .001$). 
\newline \CustomHHAIshared{HHAI--Shared} received the lowest disruptiveness ratings ($M = 2.71$, 95\% CI [2.05, 3.36]), while both \CustomHAI{HAI} ($M = 6.01$, 95\% CI [5.59, 6.42]) and \CustomHHAIpersonal{HHAI--Personal} ($M = 6.18$, 95\% CI [5.76, 6.59]) were perceived as significantly more disruptive. As illustrated in Figure~\ref{fig:used_suggestion_and_interruption}.B, post-hoc comparisons revealed that HHAI--Shared was rated as significantly less disruptive than both HAI ($p < .001$) and HHAI--Personal conditions ($p < .001$). However, HAI and HHAI--Personal did not differ significantly in disruptiveness ratings ($p = .820$). We also observed a significant order effect, where participants tended to rate the AI as more disruptive in later trials ($\beta = 0.77$, $z = 2.44$, $p = .015$).


Participants explained their differing perspectives on AI disruptiveness in Shared and Personal conditions by contrasting how Shared AI aligned with the group versus how Personal AI split attention between partners. P1 described Shared AI as integrated, noting that it \emph{``actively listen[ed] and [gave] suggestions based on our thought processes.''} (P1). Similarly, P3 observed that Personal AI encouraged side conversations, allowing her to check ideas privately before raising them with her partner. While useful individually, this parallel interaction disrupted the synchrony of the pair’s dialogue. Participants also muted the AI at times when its personal interventions interrupted ongoing talk, as one explained:  

\begin{quote}
``When neither of us had access to the same information [in Personal AI]… it’s like both of the users were using AI differently, so that’s not like working towards the shared goal.'' -- P6
\end{quote}

\subsubsection{Selective Uptake of AI Suggestions} Participants did not accept AI contributions at the same rate across conditions. A generalized linear mixed-effects model with a binomial distribution indicated significant differences across conditions (Wald $\chi^{2} = 107.2$, $p < .001$). The \CustomHAI{HAI} condition showed the highest uptake rate (83.1\%, 95\% CI [77.6, 87.4]), followed by \CustomHHAIshared{HHAI--Shared} (67.2\%, 95\% CI [59.0, 74.5]) and \CustomHHAIpersonal{HHAI--Personal} (45.0\%, 95\% CI [37.6, 52.6]). As shown in Figure~\ref{fig:used_suggestion_and_interruption}.A, all pairwise comparisons were statistically significant ($p < .05$). Experimental order did not significantly influence suggestion usage ($\beta = -0.15$, $z = -1.62$, $p = .104$).




We observed that this difference was most pronounced for proactive interventions. In HAI, participants were more likely to use proactive input, but in HHAI pairs, many of those same interventions were filtered out unless they directly supported the ongoing conversation. When proactive input fit the moment, it was valued for introducing new perspectives:  

\begin{quote}
``We would think of one way… and then AI would jump in… and we’d be like, oh, we hadn’t thought of this way before.'' -- P2
\end{quote}

Other participants emphasized the convenience or human-like quality of well-timed suggestions, as P4 described liking that he could \emph{``just glance and keep going''}, and P13 noted that \emph{``having something automatically listening felt very human-like.''} Yet participants also ignored many proactive suggestions in HHAI when they were irrelevant, treating them as noise rather than support.  

These uptake patterns add context to the disruptiveness ratings. In \CustomHAI{HAI}, where participants relied heavily on the AI, both direct and proactive contributions were more readily incorporated ($OR = 2.39, p < .001$ vs. HHAI-Shared; $OR = 6.01, p < .001$ vs. HHAI-Personal; Figure~\ref{fig:used_suggestion_and_interruption} A). In contrast, in HHAI, participants filtered suggestions more carefully, often discarding proactive input that did not align with their joint reasoning between human partners. Suggestions that fit the flow of the peer dialogue were experienced as smoother and more likely to be adopted, while irrelevant or fragmented interventions felt disruptive and were often ignored.

\revised{\subsection{Programming Performance Is Consistent Across Conditions}}

\begin{figure}[b]
    \centering
    \includegraphics[width=0.4\textwidth]{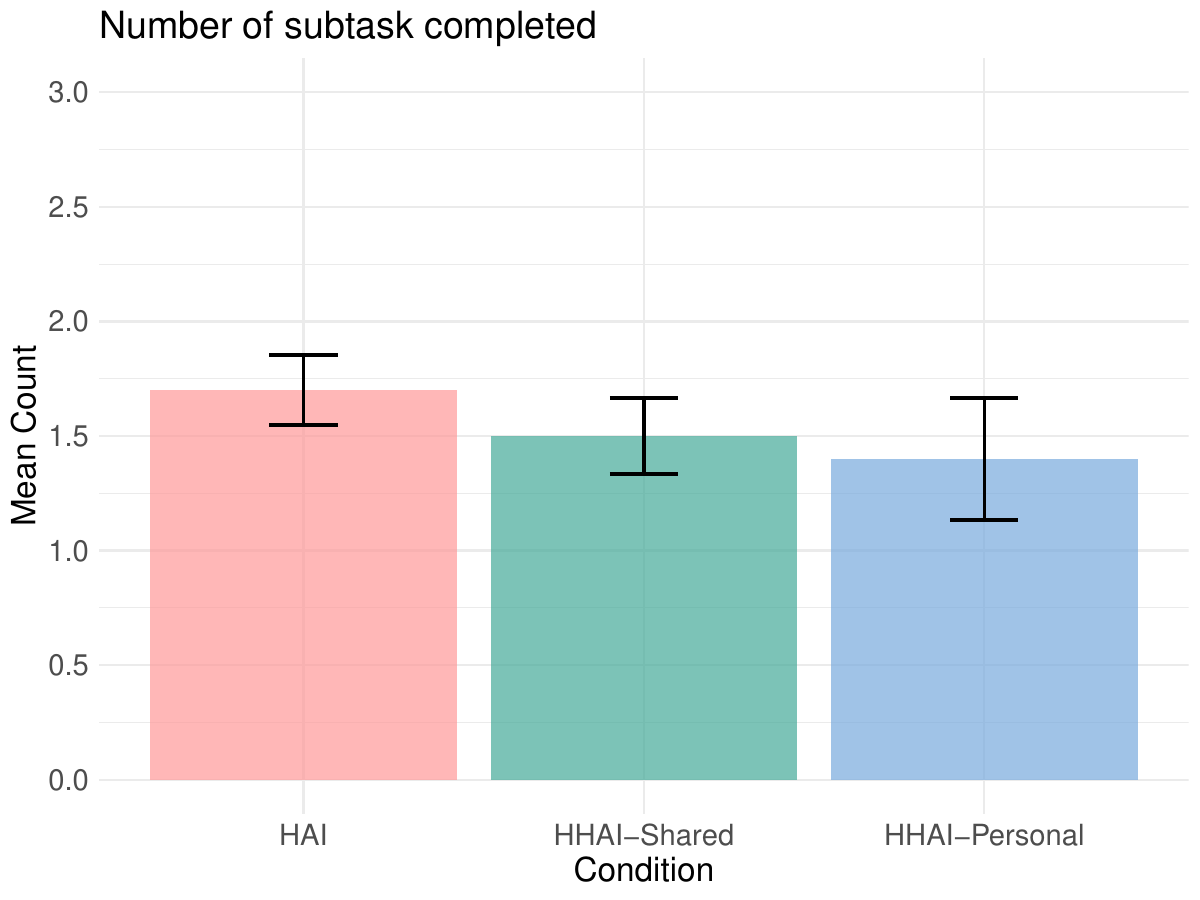}
    \caption{\revised{Mean number of subtasks completed by condition type.}}
    \label{fig:bar_chart_cnt_subtask_max3}
\end{figure}

\revised{To examine whether task completion varied across the three conditions, we fitted a Poisson Generalized Linear Mixed Model (GLMM). The analysis showed no significant differences in the number of completed subtasks across the conditions (Fig. \ref{fig:bar_chart_cnt_subtask_max3}). Relative to the \CustomHAI{HAI} baseline, neither the \CustomHHAIshared{HHAI-Shared} ($\beta = -0.11, SE = 0.35, p = 0.757$) nor the \CustomHHAIpersonal{HHAI-Personal} condition ($\beta = -0.21, SE = 0.36, p = 0.557$) differed significantly. Condition order also had no effect ($\beta = 0.17, SE = 0.18, p = 0.355$). Because session duration was held constant across conditions, task completion time was not included as an outcome measure.}



\begin{table*}[t]
\centering
\renewcommand{\arraystretch}{1}
\begin{tabular}{p{2.1cm}|p{3.3cm}|p{4cm}|p{1.2cm}|p{1.2cm}|p{1cm}}
\hline
\textbf{Category} & \textbf{Description} & \textbf{Examples} & \textbf{HHAI-S (\%)} & \textbf{HHAI-P (\%)} & \textbf{HAI (\%)} \\
\hline
\hline

Question \newline (Seek Info) 
& Asking for unknown information, opinions, or help. 
& ``What would the setup for a sliding window look like?''\newline
  ``What would I do after that?''
& \textbf{12.06} & \textbf{14.22} & \textbf{35.10} \\
\hline

Question \newline  (Seek Confirm) 
& Seeking verification or agreement. 
& ``That's what we want, right?''\newline
  ``Is that good?''
& 5.54 & 5.69 & 5.57 \\
\hline

Answer / Reply 
& Direct, informative response. 
& ``It takes the highest value.''\newline
  ``We are returning the pair.''
& 1.87 & 2.32 & 0.28 \\
\hline
\hline

Proposing \newline Strategy 
& Suggesting a high-level approach. 
& ``You wanna try using a hash set?''\newline
  ``We could sort it.''
& 4.16 & 5.31 & 3.90 \\
\hline

Proposing \newline Implementation 
& Suggesting specific code-level actions. 
& ``To get the price, we can call the previous function.''\newline
  ``You need to check for a chosen index as well.''
& 17.53 & 15.04 & 17.55 \\
\hline

Justifying \newline  Proposal 
& Explaining reasoning behind a proposal. 
& ``If best streak is bigger than the sum, it doesn't matter.''\newline
  ``That avoids recomputing values, so it's more efficient.''
& \textbf{5.34} & \textbf{2.62} & \textbf{0.00} \\
\hline
\hline

Think-Aloud  
& Verbalizing one's internal reasoning process. 
& ``I think I need to do something with sum too.''\newline
  ``Add step J to sum, then subtract start.''
& 19.89 & 23.28 & 18.11 \\
\hline

Read-Aloud 
& Reading text or code verbatim. 
& ``The problem requires analyzing daily step counts.''\newline
  ``Find the highest average step count.''
& 1.52 & 0.90 & 1.11 \\
\hline
\hline

Acknowledgment \newline / Acceptance 
& Showing understanding or agreement. 
& ``Yeah, that works, I guess.''\newline
  ``Okay, code looks good.''
& \textbf{16.42} & \textbf{14.45} & \textbf{6.69} \\
\hline

Disagreement / Rejection 
& Rejecting a statement or proposal. 
& ``But then that'd be O($n^2$).''\newline
  ``I think our approach is just… not great.''
& 2.84 & 2.77 & 1.11 \\
\hline

Coordination / Turn-Taking 
& Managing roles or sequencing actions. 
& ``Do you want to code that?''\newline
  ``I'm down for whatever.''
& 5.27 & 4.42 & 3.06 \\
\hline

Affective \newline Expression 
& Expressing emotion or social signals. 
& ``Oh, you're kidding.''\newline
  ``Oh, shoot.''
& 7.55 & 8.98 & 7.52 \\
\hline

\end{tabular}
\caption{\revised{Annotation categories, descriptions, examples, and the distribution of participant utterances across HHAI-Shared, HHAI-Personal, HAI conditions.}}
\label{tab:combined_annotation_scheme}
\end{table*}

\revised{Across all conditions, participants showed a similar pattern of progressing through the problems by completing one to two subtasks per condition (HAI: $Mean = 1.7$, $Md = 2$; HHAI-Shared: $Mean = 1.5$, $Md = 1.5$; HHAI-Personal: $Mean = 1.4$, $Md = 1$) (Fig. \ref{fig:bar_chart_cnt_subtask_max3}). Since some subtasks allowed both a brute-force and an optimized solution, we also examined the total amount of work performed, counting both completed subtasks and any optimized improvements. This combined measure showed a similarly consistent pattern across conditions (HAI: $Mean = 1.9$, $Md = 2$; HHAI-Shared: $Mean = 1.8$, $Md = 1.5$; HHAI-Personal: $Mean = 1.6$, $Md = 1$). In terms of optimizing their solutions, participants found the proactive AI suggestions were helpful, as P2 noted: \textit{"the AI would jump in and provide more efficient ways of doing the problem."} However, when the AI suggested algorithms that participants had no prior familiarity with, some chose to disregard the suggestions, as P7 explained: \textit{“sometimes the feedback was a little too advanced.”} P8 similarly commented that \textit{“the AI should know you better first… what are your current skills before giving suggestions,”} highlighting the need for AI suggestions calibrated to the user’s level of coding knowledge when introducing advanced concepts.}
\vspace{0.15cm}

\subsection{Conversational Behaviors Differ Across Conditions}

\revised{While programming performance remained stable across conditions, participants’ conversational behaviors differed markedly depending on whether they were working alone with AI or collaborating with a partner in the HHAI conditions. Overall, participants in the HHAI conditions produced substantially more conversation than those in the HAI condition, as having a peer naturally created more opportunities for verbal exchange. The average number of participant utterances was higher in both HHAI settings (HHAI-Shared: $Mean = 144.3$, $Md = 151$; HHAI-Personal: $Mean = 133.6$, $Md = 149$) compared to the individual HAI condition ($Mean = 35.9$, $Md = 35$).}

\revised{To examine how these conversational interactions differed qualitatively, we analyzed participants’ utterances and computed the distribution of utterance types across the three conditions (Table~\ref{tab:combined_annotation_scheme}). As shown in the table, the HAI condition was dominated by \textit{Seek-Info Questions} (35.10\%), many of which were short, procedural prompts such as “What do I do next?” or “What would the setup look like?” These questions were typically solution-seeking rather than reflective or explanatory, which suggests a form of problem-solving dependency on the AI. In contrast, both HHAI conditions showed higher proportions of conversation behaviors that involved confirming understanding, explaining reasoning, or coordinating next steps. For example, \textit{Acknowledgment} was considerably higher in HHAI (14.44\% to 16.42\%) compared to HAI (6.69\%), and \textit{Justifying Proposal} appeared in HHAI (2.62\% to 5.34\%) but was absent in HAI (0\%). Methodological details on our utterance analysis are provided in Appendix~\ref{Appendix:D}.}

\revised{To assess whether the pattern of utterances in Table~\ref{tab:combined_annotation_scheme} differed significantly across conditions, we conducted a Pearson Chi-square test with a Monte Carlo simulated p-value (20{,}000 permutations). The test was significant ($X^2 = 176.45$, $p < .001$), indicating that the overall pattern of conversational behaviors varied across the \CustomHAI{HAI}, \CustomHHAIshared{HHAI-Shared}, and \CustomHHAIpersonal{HHAI-Personal} conditions. We then carried out post hoc analyses of category-level proportions (Table~\ref{tab:utterance_frequencies},  Appendix~\ref{Appendix:D}) to identify which utterance types contributed most to this effect. These analyses support the descriptive pattern in Table~\ref{tab:combined_annotation_scheme}: \textit{Seek-Info Questions} are more prevalent in HAI than in HHAI, whereas conversational utterances such as \textit{Acknowledgment} and \textit{Justifying Proposal} occur significantly more often in the two HHAI conditions than in HAI.}


\section{Discussion}

\subsection{Rethinking collaborative programming with AI as augmentation of human-human collaboration}

Prior research on AI-assisted programming has predominantly focused on human--AI (HAI) dyads where an AI agent such as GitHub Copilot substitutes a human peer~\cite{ma2023ai, becker2023programming, kazemitabaar2023studying, fan2025impact, lyu2025will}. These studies consistently report faster task completion and higher code correctness, and highlight on-demand debugging and syntax assistance as concrete benefits of HAI \cite{fan2025impact}. However, participants from these studies also report a diminished sense of social presence compared with human--human settings~\cite{fan2025impact}. The substitution of a human partner reflects a broader trend in AI-supported learning where systems are primarily valued for task efficiency~\cite{roest2024next, wilhelm2025managers}. Our study challenges this substitution framing by showing that in programming practice, AI support is most valuable when it augments rather than replaces the collaborative benefits of a human partner. In both \CustomHHAIshared{Shared} and \CustomHHAIpersonal{Personal} HHAI conditions the presence of a human partner increased participants' sense of collaboration and social presence compared to the HAI setting (§5.1). Participants during HHAI settings described how it felt more natural to articulate their thinking and exchange ideas with a peer, which aligns with established accounts of the pedagogical and motivational benefits of collaborative programming~\cite{so2008student, kreijns2011measuring, vass2010peer, hostetter2013community, hanks2011pair}.  

These contrasting experiences between HAI and HHAI conditions situate our findings within the long-standing HCI discussion on augmentation versus automation \cite{licklider2008man, engelbart2023augmenting}.  Automation, or replacement, frames technology as a substitute for human roles \cite{bainbridge1983ironies, parasuraman2000model}, which is reflected in the HAI condition. \revised{For instance, in the HAI condition, participants often prompted the AI with direct solution-seeking questions, effectively shifting the sense-making steps of the problem-solving process to the AI. Augmentation, by contrast, frames technology as amplifying human agency \cite{engelbart2023augmenting, licklider2008man, grudin2002computer},which was reflected in the HHAI conditions where participants engaged in substantially more conversational activity and naturally articulated their reasoning to a partner as they worked through the problem together.} Our findings show that in collaborative programming, effective augmentation means designing AI to reinforce the collaborative mechanisms that human partners already provide. Studies in learning sciences show that a key reason another human matters is that human peers create interactional conditions ~\cite{jarvela2023human, zimmerman2011self, wu2013enhancing, vass2010peer} that AI does not. With a peer present, there is social accountability \cite{jarvela2023human} and an expectation to articulate thinking so that the partner can follow along \cite{begel2008pair}.

\subsection{Influence of Human Presence on Responsible Use of AI for Learning} 

Recent work in education shows that when learners outsource problem-solving to AI, it often reduces the reflective processes that underlie critical thinking \cite{jose2025cognitive, kosmyna2025your}. Recent studies in HCI similarly raise concerns about how overreliance on AI can lead to reduced learning engagement \cite{pu2025assistance, fan2025beware, giannakos2025promise, bai2023chatgpt}. For example, a study by \citet{fan2025beware} found that students using ChatGPT exhibited a greater tendency toward ``metacognitive laziness,'' which reduced engagement in learning and increased dependence on technology. Similarly, \citet{pu2025assistance} showed that AI programming support can cause overreliance and a lack of code understanding. Our study addresses these concerns by examining how the presence of another human partner can moderate responsible engagement with AI in collaborative programming contexts.

\textbf{Our findings show that human–human–AI triads reduced overreliance, particularly in the use of AI-generated code} (\S 5.2). This suggests that the human partners did not blindly outsource the task to the AI. Instead, the presence of a human partner encouraged them to discuss solutions and use AI suggestions more responsibly. \revised{This shift was also reflected in participants’ conversational behavior. In the HHAI conditions, participants produced more \textit{Acknowledgment} and \textit{Justifying Proposal} utterances, which reflected coordinated checks and rationalization for next-steps.} This effect became even more pronounced when shifting from personal AI to shared AI, which produced a notable increase in how participants felt accountable for understanding AI suggestions before applying them. As participants explained, simply knowing that a peer could see their interaction with the AI made them more intentional about treating AI as a learning resource rather than copy-pasting code.

This dynamic can be understood through the lens of Socially Shared Regulation of Learning (SSRL) \cite{jarvela2023human, hadwin2017self}. SSRL emphasizes how learners regulate goals and strategies collectively through mutual monitoring and accountability \cite{jarvela2023human, hadwin2017self}. In collaborative settings, regulation does not rest solely on individual cognition but develops through group interactions, where learners actively shape each other’s approaches and decisions \cite{hadwin2017self}.  In our findings, accountability was not enforced externally but was socially constructed through peer presence (\S 5.2). Knowing that a partner was observing their interaction with the AI made participants use AI more responsibly. This finding also connects to recent HCI work that explores when and whether AI outputs should remain private aids for individuals and when they should be shared as resources that peers can jointly monitor \cite{rezwana2022understanding, feng2024coprompt}. Our results contribute to this discussion by showing that shared visibility of AI suggestions can embed accountability into the interaction, making responsible use of AI in collaborative programming learning contexts more likely.

However, adding a human partner did not always automatically encourage responsible use of AI (\S 5.2). In some cases, when human-to-human collaboration was not effective (e.g., when both users communicated very little) and both participants were already used to relying heavily on AI, the presence of a human peer did little to reduce overreliance on AI. This suggests that the benefits of triadic collaboration are not automatic but depend on the quality of interaction that participants bring into the collaboration. When mutual accountability is weak, peers may reinforce each other’s reliance on AI rather than moderating it \cite{sinclair2019effects}.  Future work in collaborative learning systems in HHAI contexts should consider scaffolding productive human-to-human peer interaction so that the presence of a human partner can reliably foster more responsible engagement with AI.

\subsection{Design Implications for \revised{Human-Human-AI Triadic Programming}}
\subsubsection{Implementing Shared AI to Support Collective Learning}
Our study shows the benefits of having an AI agent shared by participants, which supports collective learning and reduces overreliance on AI (\S 5.1, \S 5.2). Having \CustomHHAIshared{shared AI} in HHAI triads also enabled participants to engage in various ways of learning. For example, some participants reported that they learned by actively explaining their thinking approach to their partners while thinking aloud (P18), with the AI providing proactive suggestions. Others noted that they learned by observing their partner’s actions (P3), particularly when the AI offered suggestions that one partner then applied in the code. Similarly, some participants asked their partners to help them better understand the AI’s suggestions (P5, P6). As such, having a shared AI was important because \textit{"see[ing] the same output of the AI is helpful, [otherwise, my partner] could be referencing something that I don't see," (P9)``}

\subsubsection{Controlled Personal and Shared Discussion}
In addition to having a shared AI, sometimes a \CustomHHAIpersonal{personal AI} can be useful when participants want to explore ideas that may not align with the current conversation. As P3 mentioned: \textit{``if I’m thinking something that may or may not be relevant, I can just use a personal AI to get a clear picture of it, and then communicate with the partner about''}. However, when participants used a personal AI, there were moments when they wanted to share what they had discussed with the AI but then needed to \textit{``describe [the personal AI suggestion] ... [and mentioning that] it could be really good if you have an option to just share [that]'' (P10)}. Hence, future work can implement a feature where users can simply click to share parts of the conversation with the AI, aligning with prior work on sharing AI suggestions \cite{feng2024coprompt}.

\subsubsection{Proactive AI Intervention to Support Collaboration}
In line with prior work that highlights the benefits of proactive AI \cite{pu2025assistance, chen2025need, houde2025controlling}, our participants also acknowledged the value of proactive AI assistance in supporting the learning process (P1, P2, P13). Compared to prior work on proactive programming assistants \cite{pu2025assistance, chen2025need}, we focused on AI proactivity in conversational interactions, since collaborative programming involves the exchange of ideas between humans \cite{kavitha2015knowledge}. Participants mentioned that having a proactive AI felt like \textit{``it was listening to us, it would jump in and provide more, like, efficient ways of doing the problem'' (P2)}. As such, this proactive AI behavior can provide \textit{``unexpected [learning] moments'' (P2)}, where participants learned something new without having to manually ask the AI. \revised{Building upon the potential of proactive AI, future work can explore more agentic AI \cite{sapkota2025ai} that can autonomously decide when to intervene, determine what type of support to provide, and adapt its behavior dynamically to the ongoing conversation or shared goals.}


\subsubsection{Pedagogically Driven AI to Facilitate Learning}
In general, participants appreciated when the AI provided just enough support for the learning process without revealing the solution too early (P11, P12, P13, P14, P18). As P12 mentioned: \textit{``It allows you to learn regardless of how you [ask] it. It would just give you hints, or the conceptual background on how you would use it ... which I thought was really good.''} One participant even suggested reducing the AI's expertise. As P9 explained: \textit{``[The AI] was really trying to give me the answer (hints), [but] I wanted to work through it myself and give it a try.''} He further described his preference of adding an option to have a \textit{``partner-based personality [AI]''} where the AI would be at the same level as the human and primarily act as a discussion partner rather than a teacher. As such, this aligns with prior work highlighting the benefits of offering multiple options to customize AI personas \cite{ha2024clochat}.

\subsubsection{Multimodal and Integrated Interface for Collaborative Programming with AI}
Participants generally appreciated the integrated interface that combined AI access with the code editor, which made it easier to access AI support. The multimodal interaction, such as simply clicking to trigger code analysis, allowed participants to get help quickly without having to manually prompt the AI. Automated voice transcription also enabled the AI to continuously listen to the conversation and proactively provide contextually relevant suggestions (P1, P2, P13). However, participants generally disliked the AI voice output and often turned it off early in the session. As P6 mentioned: \textit{``I think one problem with the AI [voice] is that if it starts to speak, it will keep on speaking ... maybe you [should be able to] barge in, and then, like, I can start speaking again.''} Even though the AI began speaking while the user was idle, once it started it would continue until finishing its sentences, which sometimes prevented participants from smoothly resuming their conversation with their partner. Hence, there is a need for a better turn-taking mechanism that adapts to the flow of conversation \cite{skantze2025applying}.

\section{Limitations}
\revised{We first acknowledge the limitation of not having a human–human collaborative programming baseline (no-AI). Although our findings suggest that the AI supported collaboration in meaningful ways, the absence of a human–human baseline limits our ability to determine the extent of that contribution. Without a no-AI comparison, we cannot fully assess how much the AI improved collaboration beyond what two humans might achieve on their own, or whether it may have disrupted the collaboration instead.} 

\revised{Secondly, our study involved college students from a single institution working on LeetCode-style problems. While this allows us to explore initial findings in a controlled setting, it limits how far the results can generalize to other programming contexts or learner populations. The single-session format also limits us from understanding longer-term development or how triadic collaboration might evolve over time. Future work could conduct longitudinal studies and explore a broader range of programming tasks to better capture how triadic collaboration unfolds across different contexts and over extended periods. It could also examine different populations, such as professional software engineers, to understand how these dynamics operate in more varied and professional settings.}

\revised{Thirdly, all participant pairs consisted of students who already knew their partner and registered together. While we did not measure the closeness of these relationships, participants were asked to recruit someone they knew, so pairs were not strangers and were not randomly assigned. This familiarity may have influenced how they collaborated, as prior work shows that social relationships can shape collaborative programming dynamics \cite{villamor2018friends, zhong2016impact}. As a result, our findings primarily reflect triadic collaboration among partners with some prior familiarity. Future work can examine how these dynamics play out among pairs who have little or no prior relationship.}

Lastly, although the experimental design used full counterbalancing, three consistent order effects emerged across conditions. As sessions progressed, participants increasingly relied on AI-generated code, reported less responsibility for understanding AI suggestions, and perceived the AI as more disruptive. Because counterbalancing ensured that order was orthogonal to experimental condition, these trends do not confound the primary comparisons across conditions. These effects can be interpreted as general patterns of how user behavior and perception evolve during a prolonged, 90-minute programming session with an AI assistant. Such patterns may reflect a combination of adaptation and fatigue effects. Early in the session, participants may have been more cautious, carefully checking AI output and assuming greater responsibility for integrating suggestions. Over time, however, familiarity with the tool and cognitive fatigue may have fostered greater reliance on AI code, accompanied by reduced vigilance and heightened perceptions of the AI as interruptive. These dynamics highlight that the temporal trajectory of interaction can shape both behavioral reliance on AI and subjective evaluations of its role, independent of the specific interaction scenario. 


\section{Conclusion}
Our study explored \revised{human-human-AI triadic programming}, where an
AI agent serves as an additional collaborator rather than a substitute for a human partner. Through a study with 20 participants, we
show that this triadic collaboration can enhance collaborative learning and social presence compared to a human–AI baseline,  \revised{while maintaining similar programming performance}. Our findings also highlight how human partners encourage more accountable AI use, primarily reducing reliance on AI-generated code.
This effect becomes amplified when shifting from personal AI to shared AI support, which further increases participants’ sense of
responsibility to understand AI suggestions before applying them. Taken together, our findings suggest that the value of AI in educational settings extends beyond efficiency gains. By embedding AI within peer collaboration, systems can strengthen the social and pedagogical dynamics that underpin effective learning. Building on these insights, we propose design implications for future systems: make AI outputs visible to peers to scaffold accountability, align proactive interventions with conversational flow, and reinforce the social and pedagogical benefits of working with others.

\bibliographystyle{ACM-Reference-Format}
\bibliography{sample-base}

@String{Computing = "Computing" }

@String{Computer = "{IEEE} Computer" }

@String{Academic = "Academic Press" }

@String{Springer = "Springer-Verlag" }

@article{hanks2011pair,
  title={Pair programming in education: A literature review},
  author={Hanks, Brian and Fitzgerald, Sue and McCauley, Ren{\'e}e and Murphy, Laurie and Zander, Carol},
  journal={Computer Science Education},
  volume={21},
  number={2},
  pages={135--173},
  year={2011},
  publisher={Taylor \& Francis}
}

@article{bryant2008pair,
  title={Pair programming and the mysterious role of the navigator},
  author={Bryant, Sallyann and Romero, Pablo and du Boulay, Benedict},
  journal={International Journal of Human-Computer Studies},
  volume={66},
  number={7},
  pages={519--529},
  year={2008},
  publisher={Elsevier}
}

@article{williams2001support,
  title={In support of student pair-programming},
  author={Williams, Laurie and Upchurch, Richard L},
  journal={ACM Sigcse Bulletin},
  volume={33},
  number={1},
  pages={327--331},
  year={2001},
  publisher={ACM New York, NY, USA}
}

@inproceedings{hulkko2005multiple,
  title={A multiple case study on the impact of pair programming on product quality},
  author={Hulkko, Hanna and Abrahamsson, Pekka},
  booktitle={Proceedings of the 27th international conference on Software engineering},
  pages={495--504},
  year={2005}
}

@inproceedings{celepkolu2018importance,
  title={The importance of producing shared code through pair programming},
  author={Celepkolu, Mehmet and Boyer, Kristy Elizabeth},
  booktitle={Proceedings of the 49th ACM technical symposium on computer science education},
  pages={765--770},
  year={2018}
}

@article{ma2023ai,
  title={Is ai the better programming partner? human-human pair programming vs. human-ai pair programming},
  author={Ma, Qianou and Wu, Tongshuang and Koedinger, Kenneth},
  journal={arXiv preprint arXiv:2306.05153},
  year={2023}
}

@article{fan2025impact,
  title={The impact of AI-assisted pair programming on student motivation, programming anxiety, collaborative learning, and programming performance: a comparative study with traditional pair programming and individual approaches},
  author={Fan, Guangrui and Liu, Dandan and Zhang, Rui and Pan, Lihu},
  journal={International Journal of STEM Education},
  volume={12},
  number={1},
  pages={16},
  year={2025},
  publisher={Springer}
}

@article{beck2013cooperative,
  title={Cooperative learning instructional methods for CS1: Design, implementation, and evaluation},
  author={Beck, Leland and Chizhik, Alexander},
  journal={ACM Transactions on Computing Education (TOCE)},
  volume={13},
  number={3},
  pages={1--21},
  year={2013},
  publisher={ACM New York, NY, USA}
}

@article{daryanto2025designing,
  title={Designing Conversational AI to Support Think-Aloud Practice in Technical Interview Preparation for CS Students},
  author={Daryanto, Taufiq and Stil, Sophia and Ding, Xiaohan and Manesh, Daniel and Lee, Sang Won and Lee, Tim and Lunn, Stephanie and Rodriguez, Sarah and Brown, Chris and Rho, Eugenia},
  journal={arXiv preprint arXiv:2507.14418},
  year={2025}
}

@article{wu2013enhancing,
  title={Enhancing motivation and engagement through collaborative discussion.},
  author={Wu, Xiaoying and Anderson, Richard C and Nguyen-Jahiel, Kim and Miller, Brian},
  journal={Journal of Educational Psychology},
  volume={105},
  number={3},
  pages={622},
  year={2013},
  publisher={American Psychological Association}
}

@inproceedings{lyu2025will,
  title={Will Your Next Pair Programming Partner Be Human? An Empirical Evaluation of Generative AI as a Collaborative Teammate in a Semester-Long Classroom Setting},
  author={Lyu, Wenhan and Wang, Yimeng and Sun, Yifan and Zhang, Yixuan},
  booktitle={Proceedings of the Twelfth ACM Conference on Learning@ Scale},
  pages={83--94},
  year={2025}
}

@inproceedings{pu2025assistance,
  title={Assistance or disruption? exploring and evaluating the design and trade-offs of proactive ai programming support},
  author={Pu, Kevin and Lazaro, Daniel and Arawjo, Ian and Xia, Haijun and Xiao, Ziang and Grossman, Tovi and Chen, Yan},
  booktitle={Proceedings of the 2025 CHI Conference on Human Factors in Computing Systems},
  pages={1--21},
  year={2025}
}

@article{gu2021predicting,
  title={Predicting the future of artificial intelligence and its educational impact: A thought experiment based on social science fiction},
  author={Gu, X and Cai, H},
  journal={Educational research},
  volume={42},
  number={05},
  pages={137--147},
  year={2021}
}

@article{ayeni2024ai,
  title={AI in education: A review of personalized learning and educational technology},
  author={Ayeni, Oyebola Olusola and Al Hamad, Nancy Mohd and Chisom, Onyebuchi Nneamaka and Osawaru, Blessing and Adewusi, Ololade Elizabeth},
  journal={GSC Advanced Research and Reviews},
  volume={18},
  number={2},
  pages={261--271},
  year={2024}
}

@inproceedings{phung2023generative,
  title={Generative AI for programming education: benchmarking ChatGPT, GPT-4, and human tutors},
  author={Phung, Tung and P{\u{a}}durean, Victor-Alexandru and Cambronero, Jos{\'e} and Gulwani, Sumit and Kohn, Tobias and Majumdar, Rupak and Singla, Adish and Soares, Gustavo},
  booktitle={Proceedings of the 2023 ACM Conference on International Computing Education Research-Volume 2},
  pages={41--42},
  year={2023}
}

@inproceedings{chen2025need,
  title={Need help? designing proactive ai assistants for programming},
  author={Chen, Valerie and Zhu, Alan and Zhao, Sebastian and Mozannar, Hussein and Sontag, David and Talwalkar, Ameet},
  booktitle={Proceedings of the 2025 CHI Conference on Human Factors in Computing Systems},
  pages={1--18},
  year={2025}
}

@inproceedings{wang2023exploring,
  title={Exploring the Role of AI Assistants in Computer Science Education: Methods, Implications, and Instructor Perspectives},
  author={Wang, Tianjia and D{\'\i}az, Daniel Vargas and Brown, Chris and Chen, Yan},
  booktitle={2023 IEEE Symposium on Visual Languages and Human-Centric Computing (VL/HCC)},
  pages={92--102},
  year={2023},
  organization={IEEE}
}

@inproceedings{ghimire2024coding,
  title={Coding with ai: How are tools like chatgpt being used by students in foundational programming courses},
  author={Ghimire, Aashish and Edwards, John},
  booktitle={International Conference on Artificial Intelligence in Education},
  pages={259--267},
  year={2024},
  organization={Springer}
}

@inproceedings{sarsa2022automatic,
  title={Automatic generation of programming exercises and code explanations using large language models},
  author={Sarsa, Sami and Denny, Paul and Hellas, Arto and Leinonen, Juho},
  booktitle={Proceedings of the 2022 ACM Conference on International Computing Education Research-Volume 1},
  pages={27--43},
  year={2022}
}

@inproceedings{ma2024teach,
  title={How to teach programming in the ai era? using llms as a teachable agent for debugging},
  author={Ma, Qianou and Shen, Hua and Koedinger, Kenneth and Wu, Sherry Tongshuang},
  booktitle={International Conference on Artificial Intelligence in Education},
  pages={265--279},
  year={2024},
  organization={Springer}
}

@article{ma2024you,
  title={What you say= what you want? Teaching humans to articulate requirements for LLMs},
  author={Ma, Qianou and Peng, Weirui and Shen, Hua and Koedinger, Kenneth and Wu, Tongshuang},
  journal={arXiv preprint arXiv:2409.08775},
  year={2024}
}

@article{guo2023six,
  title={Six Opportunities for scientists and engineers to learn programming using AI Tools such as ChatGPT},
  author={Guo, Philip J},
  journal={Computing in Science \& Engineering},
  volume={25},
  number={3},
  pages={73--78},
  year={2023},
  publisher={IEEE}
}

@inproceedings{jin2024teach,
  title={Teach AI How to Code: Using Large Language Models as Teachable Agents for Programming Education},
  author={Jin, Hyoungwook and Lee, Seonghee and Shin, Hyungyu and Kim, Juho},
  booktitle={Proceedings of the CHI Conference on Human Factors in Computing Systems},
  pages={1--28},
  year={2024}
}

@misc{mdn_websockets,
  author       = {{MDN Web Docs}},
  title        = {WebSockets API},
  year         = {2024},
  howpublished = {\url{https://developer.mozilla.org/en-US/docs/Web/API/WebSockets_API}},
  note         = {Accessed: 2025-08-19}
}

@misc{leetcode,
  author       = {{LeetCode}},
  title        = {LeetCode},
  year         = {2025},
  howpublished = {\url{http://leetcode.com/}},
  note         = {Accessed: 2025-08-19}
}

@misc{vscode_liveshare,
  author       = {{Microsoft}},
  title        = {Visual Studio Live Share},
  year         = {2025},
  howpublished = {\url{https://visualstudio.microsoft.com/services/live-share/}},
  note         = {Accessed: 2025-08-19}
}

@inproceedings{roest2024next,
  title={Next-step hint generation for introductory programming using large language models},
  author={Roest, Lianne and Keuning, Hieke and Jeuring, Johan},
  booktitle={Proceedings of the 26th Australasian Computing Education Conference},
  pages={144--153},
  year={2024}
}

@inproceedings{houde2025controlling,
  title={Controlling AI Agent Participation in Group Conversations: A Human-Centered Approach},
  author={Houde, Stephanie and Brimijoin, Kristina and Muller, Michael and Ross, Steven I and Silva Moran, Dario Andres and Gonzalez, Gabriel Enrique and Kunde, Siya and Foreman, Morgan A and Weisz, Justin D},
  booktitle={Proceedings of the 30th International Conference on Intelligent User Interfaces},
  pages={390--408},
  year={2025}
}

@article{collins2024building,
  title={Building machines that learn and think with people},
  author={Collins, Katherine M and Sucholutsky, Ilia and Bhatt, Umang and Chandra, Kartik and Wong, Lionel and Lee, Mina and Zhang, Cedegao E and Zhi-Xuan, Tan and Ho, Mark and Mansinghka, Vikash and others},
  journal={Nature human behaviour},
  volume={8},
  number={10},
  pages={1851--1863},
  year={2024},
  publisher={Nature Publishing Group UK London}
}

@article{zakaria2022two,
  title={Two-computer pair programming: Exploring a feedback intervention to improve collaborative talk in elementary students},
  author={Zakaria, Zarifa and Vandenberg, Jessica and Tsan, Jennifer and Boulden, Danielle Cadieux and Lynch, Collin F and Boyer, Kristy Elizabeth and Wiebe, Eric N},
  journal={Computer Science Education},
  volume={32},
  number={1},
  pages={3--29},
  year={2022},
  publisher={Taylor \& Francis}
}

@inproceedings{gonzalez2024collaborative,
  title={Collaborative Canvas: A Tool for Exploring LLM Use in Group Ideation Tasks.},
  author={Gonzalez, Gabriel Enrique and Moran, Dario Andres Silva and Houde, Stephanie and He, Jessica and Ross, Steven I and Muller, Michael J and Kunde, Siya and Weisz, Justin D},
  booktitle={IUI Workshops},
  year={2024}
}

@inproceedings{hughes2020remote,
  title={Remote pair programming},
  author={Hughes, Janet and Walshe, Ann and Law, Bobby and Murphy, Brendan},
  booktitle={12th International Conference on Computer Supported Education},
  pages={476--483},
  year={2020},
  organization={SciTePress}
}

@inproceedings{gupta2025multimodal,
  title={Multimodal Programming in Computer Science with Interactive Assistance Powered by Large Language Model},
  author={Gupta, Rajan Das and Hosain, Md Tanzib and Mridha, Muhammad Firoz and Ahmed, Salah Uddin},
  booktitle={International Conference on Human-Computer Interaction},
  pages={59--69},
  year={2025},
  organization={Springer}
}

@online{CodeMirror,
  author = {Marijn Haverbeke},
  title  = {CodeMirror},
  year   = {2007--2017},
  url    = {https://codemirror.net/}
}

@inproceedings{singh2024figura11y,
  title={Figura11y: Ai assistance for writing scientific alt text},
  author={Singh, Nikhil and Wang, Lucy Lu and Bragg, Jonathan},
  booktitle={Proceedings of the 29th International Conference on Intelligent User Interfaces},
  pages={886--906},
  year={2024}
}

@inproceedings{zhang2023visar,
  title={Visar: A human-ai argumentative writing assistant with visual programming and rapid draft prototyping},
  author={Zhang, Zheng and Gao, Jie and Dhaliwal, Ranjodh Singh and Li, Toby Jia-Jun},
  booktitle={Proceedings of the 36th annual ACM symposium on user interface software and technology},
  pages={1--30},
  year={2023}
}

@article{desolda2025understanding,
  title={Understanding User Mental Models in AI-Driven Code Completion Tools: Insights from an Elicitation Study},
  author={Desolda, Giuseppe and Esposito, Andrea and Greco, Francesco and Tucci, Cesare and Buono, Paolo and Piccinno, Antonio},
  journal={arXiv preprint arXiv:2502.02194},
  year={2025}
}

@article{sun2025don,
  title={Don’t complete it! Preventing unhelpful code completion for productive and sustainable neural code completion systems},
  author={Sun, Zhensu and Du, Xiaoning and Song, Fu and Wang, Shangwen and Ni, Mingze and Li, Li and Lo, David},
  journal={ACM Transactions on Software Engineering and Methodology},
  volume={34},
  number={1},
  pages={1--22},
  year={2025},
  publisher={ACM New York, NY}
}

@inproceedings{wang2020human,
  title={From human-human collaboration to Human-AI collaboration: Designing AI systems that can work together with people},
  author={Wang, Dakuo and Churchill, Elizabeth and Maes, Pattie and Fan, Xiangmin and Shneiderman, Ben and Shi, Yuanchun and Wang, Qianying},
  booktitle={Extended abstracts of the 2020 CHI conference on human factors in computing systems},
  pages={1--6},
  year={2020}
}

@article{daryanto2025conversate,
  title={Conversate: Supporting reflective learning in interview practice through interactive simulation and dialogic feedback},
  author={Daryanto, Taufiq and Ding, Xiaohan and Wilhelm, Lance T and Stil, Sophia and Knutsen, Kirk McInnis and Rho, Eugenia H},
  journal={Proceedings of the ACM on Human-Computer Interaction},
  volume={9},
  number={1},
  pages={1--32},
  year={2025},
  publisher={ACM New York, NY, USA}
}

@article{chen2022human,
  title={Human-AI cooperation in education: human in loop and teaching as leadership},
  author={Chen, Feng},
  journal={Journal of Educational Technology and Innovation},
  volume={2},
  number={1},
  year={2022}
}

@article{wang2024investigating,
  title={Investigating dialogic interaction in K12 online one-on-one mathematics tutoring using AI and sequence mining techniques},
  author={Wang, Deliang and Shan, Dapeng and Ju, Ran and Kao, Ben and Zhang, Chenwei and Chen, Gaowei},
  journal={Education and Information Technologies},
  pages={1--26},
  year={2024},
  publisher={Springer}
}

@article{fahnenstich2024trusting,
  title={Trusting under risk--comparing human to AI decision support agents},
  author={Fahnenstich, Hannah and Rieger, Tobias and Roesler, Eileen},
  journal={Computers in Human Behavior},
  volume={153},
  pages={108107},
  year={2024},
  publisher={Elsevier}
}

@article{wilhelm2025managers,
  title={How Managers Perceive AI-Assisted Conversational Training for Workplace Communication},
  author={Wilhelm, Lance T and Ding, Xiaohan and Knutsen, Kirk McInnis and Carik, Buse and Rho, Eugenia H},
  journal={arXiv preprint arXiv:2505.14452},
  year={2025}
}

@article{cockrell2000context,
  title={A context for learning: Collaborative groups in the problem-based learning environment},
  author={Cockrell, Karen Sunday and Caplow, Julie A Hughes and Donaldson, Joe F},
  journal={The Review of Higher Education},
  volume={23},
  number={3},
  pages={347--363},
  year={2000},
  publisher={Johns Hopkins University Press}
}

@article{javed2025exploring,
  title={Exploring how AI can be used to Promote Collaboration in group Project reduce Conflict in Team Dynamics and Enhance Cooperative Learning Experiences},
  author={Javed, Umer and Rohilla, Atiya and Adnan, Ghazal and Taj, Naveen},
  journal={Review of Applied Management and Social Sciences},
  volume={8},
  number={1},
  pages={129--141},
  year={2025}
}

@inproceedings{muller2024group,
  title={Group brainstorming with an ai agent: Creating and selecting ideas},
  author={Muller, Michael and Houde, Stephanie and Gonzalez, Gabriel and Brimijoin, Kristina and Ross, Steven I and Moran, Dario Andres Silva and Weisz, Justin D},
  booktitle={International conference on computational creativity},
  pages={10},
  year={2024}
}

@article{de2014team,
  title={Team decision making and individual satisfaction with the team},
  author={De la Torre-Ruiz, Jos{\'e} M and Ferr{\'o}n-V{\'\i}lchez, Vera and Ortiz-de-Mandojana, Natalia},
  journal={Small group research},
  volume={45},
  number={2},
  pages={198--216},
  year={2014},
  publisher={Sage Publications Sage CA: Los Angeles, CA}
}

@inproceedings{chiang2024enhancing,
  title={Enhancing ai-assisted group decision making through llm-powered devil's advocate},
  author={Chiang, Chun-Wei and Lu, Zhuoran and Li, Zhuoyan and Yin, Ming},
  booktitle={Proceedings of the 29th International Conference on Intelligent User Interfaces},
  pages={103--119},
  year={2024}
}

@misc{github_copilot,
  author       = {GitHub},
  title        = {GitHub Copilot},
  year         = {2025},
  url          = {https://github.com/features/copilot},
  note         = {Accessed: 2025-09-10}
}

@inproceedings{kazemitabaar2023studying,
  title={Studying the effect of AI code generators on supporting novice learners in introductory programming},
  author={Kazemitabaar, Majeed and Chow, Justin and Ma, Carl Ka To and Ericson, Barbara J and Weintrop, David and Grossman, Tovi},
  booktitle={Proceedings of the 2023 CHI conference on human factors in computing systems},
  pages={1--23},
  year={2023}
}

@inproceedings{becker2023programming,
  title={Programming is hard-or at least it used to be: Educational opportunities and challenges of ai code generation},
  author={Becker, Brett A and Denny, Paul and Finnie-Ansley, James and Luxton-Reilly, Andrew and Prather, James and Santos, Eddie Antonio},
  booktitle={Proceedings of the 54th ACM Technical Symposium on Computer Science Education V. 1},
  pages={500--506},
  year={2023}
}

@article{fan2025beware,
  title={Beware of metacognitive laziness: Effects of generative artificial intelligence on learning motivation, processes, and performance},
  author={Fan, Yizhou and Tang, Luzhen and Le, Huixiao and Shen, Kejie and Tan, Shufang and Zhao, Yueying and Shen, Yuan and Li, Xinyu and Ga{\v{s}}evi{\'c}, Dragan},
  journal={British Journal of Educational Technology},
  volume={56},
  number={2},
  pages={489--530},
  year={2025},
  publisher={Wiley Online Library}
}

@article{giannakos2025promise,
  title={The promise and challenges of generative AI in education},
  author={Giannakos, Michail and Azevedo, Roger and Brusilovsky, Peter and Cukurova, Mutlu and Dimitriadis, Yannis and Hernandez-Leo, Davinia and J{\"a}rvel{\"a}, Sanna and Mavrikis, Manolis and Rienties, Bart},
  journal={Behaviour \& Information Technology},
  volume={44},
  number={11},
  pages={2518--2544},
  year={2025},
  publisher={Taylor \& Francis}
}

@article{bai2023chatgpt,
  title={ChatGPT: The cognitive effects on learning and memory},
  author={Bai, Long and Liu, Xiangfei and Su, Jiacan},
  journal={Brain-X},
  volume={1},
  number={3},
  pages={e30},
  year={2023},
  publisher={Wiley Online Library}
}

@article{jarvela2023human,
  title={Human and artificial intelligence collaboration for socially shared regulation in learning},
  author={J{\"a}rvel{\"a}, Sanna and Nguyen, Andy and Hadwin, Allyson},
  journal={British Journal of Educational Technology},
  volume={54},
  number={5},
  pages={1057--1076},
  year={2023},
  publisher={Wiley Online Library}
}

@article{zimmerman2011self,
  title={Self-regulated learning and performance: An introduction and an overview},
  author={Zimmerman, Barry J and Schunk, Dale H},
  journal={Handbook of self-regulation of learning and performance},
  pages={15--26},
  year={2011},
  publisher={Routledge}
}

@article{vass2010peer,
  title={Peer collaboration and learning in the classroom},
  author={Vass, Eva and Littleton, Karen},
  journal={International handbook of psychology in education},
  pages={105--135},
  year={2010},
  publisher={Emerald Leeds,, UK}
}

@article{hostetter2013community,
  title={Community matters: Social presence and learning outcomes.},
  author={Hostetter, Carol},
  journal={Journal of the Scholarship of Teaching and Learning},
  volume={13},
  number={1},
  pages={77--86},
  year={2013},
  publisher={ERIC}
}

@article{so2008student,
  title={Student perceptions of collaborative learning, social presence and satisfaction in a blended learning environment: Relationships and critical factors},
  author={So, Hyo-Jeong and Brush, Thomas A},
  journal={Computers \& education},
  volume={51},
  number={1},
  pages={318--336},
  year={2008},
  publisher={Elsevier}
}

@article{kreijns2011measuring,
  title={Measuring perceived social presence in distributed learning groups},
  author={Kreijns, Karel and Kirschner, Paul A and Jochems, Wim and Van Buuren, Hans},
  journal={Education and Information Technologies},
  volume={16},
  number={4},
  pages={365--381},
  year={2011},
  publisher={Springer}
}

@article{braun2021thematic,
  title={Thematic analysis: A practical guide},
  author={Braun, Virginia and Clarke, Victoria},
  year={2021},
  publisher={SAGE publications Ltd}
}

@incollection{hadwin2017self,
  title={Self-regulation, co-regulation, and shared regulation in collaborative learning environments},
  author={Hadwin, Allyson and J{\"a}rvel{\"a}, Sanna and Miller, Mariel},
  booktitle={Handbook of self-regulation of learning and performance},
  pages={83--106},
  year={2017},
  publisher={Routledge}
}

@inproceedings{feng2024coprompt,
  title={Coprompt: Supporting prompt sharing and referring in collaborative natural language programming},
  author={Feng, Li and Yen, Ryan and You, Yuzhe and Fan, Mingming and Zhao, Jian and Lu, Zhicong},
  booktitle={Proceedings of the 2024 CHI Conference on Human Factors in Computing Systems},
  pages={1--21},
  year={2024}
}

@article{kavitha2015knowledge,
  title={Knowledge sharing through pair programming in learning environments: An empirical study},
  author={Kavitha, RK and Ahmed, MS Irfan},
  journal={Education and Information Technologies},
  volume={20},
  number={2},
  pages={319--333},
  year={2015},
  publisher={Springer}
}

@inproceedings{ha2024clochat,
  title={CloChat: Understanding how people customize, interact, and experience personas in large language models},
  author={Ha, Juhye and Jeon, Hyeon and Han, Daeun and Seo, Jinwook and Oh, Changhoon},
  booktitle={Proceedings of the 2024 CHI Conference on Human Factors in Computing Systems},
  pages={1--24},
  year={2024}
}

@inproceedings{skantze2025applying,
  title={Applying general turn-taking models to conversational human-robot interaction},
  author={Skantze, Gabriel and Irfan, Bahar},
  booktitle={2025 20th ACM/IEEE International Conference on Human-Robot Interaction (HRI)},
  pages={859--868},
  year={2025},
  organization={IEEE}
}

@article{sinclair2019effects,
  title={The effects of peer-assisted learning on disruptive behavior and academic engagement},
  author={Sinclair, Anne C and Gesel, Samantha A and Lemons, Christopher J},
  journal={Journal of Positive Behavior Interventions},
  volume={21},
  number={4},
  pages={238--248},
  year={2019},
  publisher={SAGE Publications Sage CA: Los Angeles, CA}
}

@inproceedings{mowar2025codea11y,
  title={CodeA11y: Making AI Coding Assistants Useful for Accessible Web Development},
  author={Mowar, Peya and Peng, Yi-Hao and Wu, Jason and Steinfeld, Aaron and Bigham, Jeffrey P},
  booktitle={Proceedings of the 2025 CHI Conference on Human Factors in Computing Systems},
  pages={1--15},
  year={2025}
}

@inproceedings{zhao2025codinggenie,
  title={Codinggenie: A proactive llm-powered programming assistant},
  author={Zhao, Sebastian and Zhu, Alan and Mozannar, Hussein and Sontag, David and Talwalkar, Ameet and Chen, Valerie},
  booktitle={Proceedings of the 33rd ACM International Conference on the Foundations of Software Engineering},
  pages={1168--1172},
  year={2025}
}

@article{to2016making,
  title={Making productive use of exemplars: Peer discussion and teacher guidance for positive transfer of strategies},
  author={To, Jessica and Carless, David},
  journal={Journal of Further and Higher Education},
  volume={40},
  number={6},
  pages={746--764},
  year={2016},
  publisher={Taylor \& Francis}
}

@article{baidoo2023education,
  title={Education in the era of generative artificial intelligence (AI): Understanding the potential benefits of ChatGPT in promoting teaching and learning},
  author={Baidoo-Anu, David and Ansah, Leticia Owusu},
  journal={Journal of AI},
  volume={7},
  number={1},
  pages={52--62},
  year={2023},
  publisher={{\.I}zmir Academy Association}
}

@article{de2024learning,
  title={A learning analytics-based collaborative conversational agent to foster productive dialogue in inquiry learning},
  author={de Araujo, Adelson and Papadopoulos, Pantelis M and McKenney, Susan and de Jong, Ton},
  journal={Journal of computer assisted learning},
  volume={40},
  number={6},
  pages={2700--2714},
  year={2024},
  publisher={Wiley Online Library}
}

@article{edmondson2021reflections,
  title={Reflections: voice and silence in workplace conversations},
  author={Edmondson, Amy C and Besieux, Tijs},
  journal={Journal of Change Management},
  volume={21},
  number={3},
  pages={269--286},
  year={2021},
  publisher={Taylor \& Francis}
}

@article{haindl2024students,
  title={Students’ experiences of using ChatGPT in an undergraduate programming course},
  author={Haindl, Philipp and Weinberger, Gerald},
  journal={IEEE Access},
  volume={12},
  pages={43519--43529},
  year={2024},
  publisher={IEEE}
}

@inproceedings{lin2021pdl,
  title={PDL: scaffolding problem solving in programming courses},
  author={Lin, Shu and Meng, Na and Kafura, Dennis and Li, Wenxin},
  booktitle={Proceedings of the 26th ACM Conference on Innovation and Technology in Computer Science Education V. 1},
  pages={185--191},
  year={2021}
}

@inproceedings{he2024ai,
  title={AI and the Future of Collaborative Work: Group Ideation with an LLM in a Virtual Canvas},
  author={He, Jessica and Houde, Stephanie and Gonzalez, Gabriel E and Silva Moran, Dar{\'\i}o Andr{\'e}s and Ross, Steven I and Muller, Michael and Weisz, Justin D},
  booktitle={Proceedings of the 3rd Annual Meeting of the Symposium on Human-Computer Interaction for Work},
  pages={1--14},
  year={2024}
}

@inproceedings{zhang2025ladica,
  title={LADICA: a large shared display interface for generative AI cognitive assistance in co-located team collaboration},
  author={Zhang, Zheng and Peng, Weirui and Chen, Xinyue and Cao, Luke and Li, Toby Jia-Jun},
  booktitle={Proceedings of the 2025 CHI Conference on Human Factors in Computing Systems},
  pages={1--22},
  year={2025}
}

@article{zhu2024exploring,
  title={Exploring the impact of ChatGPT on art creation and collaboration: Benefits, challenges and ethical implications},
  author={Zhu, Sijin and Wang, Zheng and Zhuang, Yuan and Jiang, Yuyang and Guo, Mengyao and Zhang, Xiaolin and Gao, Ze},
  journal={Telematics and Informatics Reports},
  volume={14},
  pages={100138},
  year={2024},
  publisher={Elsevier}
}

@article{fahad2024role,
  title={The role of ChatGpt in knowledge sharing and collaboration within digital workplaces: a systematic review},
  author={Fahad, Sheikh Abdulaziz and Salloum, Said A and Shaalan, Khaled},
  journal={Artificial intelligence in education: The power and dangers of ChatGPT in the classroom},
  pages={259--282},
  year={2024},
  publisher={Springer}
}

@inproceedings{dominic2020remote,
  title={Remote pair programming in virtual reality},
  author={Dominic, James and Tubre, Brock and Ritter, Charles and Houser, Jada and Smith, Colton and Rodeghero, Paige},
  booktitle={2020 IEEE international conference on software maintenance and evolution (ICSME)},
  pages={406--417},
  year={2020},
  organization={IEEE}
}

@misc{openai2025chatgpt,
  author       = {OpenAI},
  title        = {ChatGPT (GPT-5)},
  year         = {2025},
  howpublished = {\url{https://chat.openai.com/}},
  note         = {ChatGPT}
}

@article{jose2025cognitive,
  title={The cognitive paradox of AI in education: between enhancement and erosion},
  author={Jose, Binny and Cherian, Jaya and Verghis, Alie Molly and Varghise, Sony Mary and S, Mumthas and Joseph, Sibichan},
  journal={Frontiers in Psychology},
  volume={16},
  pages={1550621},
  year={2025},
  publisher={Frontiers Media SA}
}

@article{kosmyna2025your,
  title={Your brain on chatgpt: Accumulation of cognitive debt when using an ai assistant for essay writing task},
  author={Kosmyna, Nataliya and Hauptmann, Eugene and Yuan, Ye Tong and Situ, Jessica and Liao, Xian-Hao and Beresnitzky, Ashly Vivian and Braunstein, Iris and Maes, Pattie},
  journal={arXiv preprint arXiv:2506.08872},
  volume={4},
  year={2025}
}

@inproceedings{rezwana2022understanding,
  title={Understanding user perceptions, collaborative experience and user engagement in different human-AI interaction designs for co-creative systems},
  author={Rezwana, Jeba and Maher, Mary Lou},
  booktitle={Proceedings of the 14th Conference on Creativity and Cognition},
  pages={38--48},
  year={2022}
}

@article{licklider2008man,
  title={Man-computer symbiosis},
  author={Licklider, Joseph CR},
  journal={IRE transactions on human factors in electronics},
  number={1},
  pages={4--11},
  year={2008},
  publisher={IEEE}
}

@incollection{engelbart2023augmenting,
  title={Augmenting human intellect: A conceptual framework},
  author={Engelbart, Douglas C},
  booktitle={Augmented education in the global age},
  pages={13--29},
  year={2023},
  publisher={Routledge}
}

@incollection{bainbridge1983ironies,
  title={Ironies of automation},
  author={Bainbridge, Lisanne},
  booktitle={Analysis, design and evaluation of man--machine systems},
  pages={775--779},
  year={1983},
  publisher={Elsevier}
}

@article{parasuraman2000model,
  title={A model for types and levels of human interaction with automation},
  author={Parasuraman, Raja and Sheridan, Thomas B and Wickens, Christopher D},
  journal={IEEE Transactions on systems, man, and cybernetics-Part A: Systems and Humans},
  volume={30},
  number={3},
  pages={286--297},
  year={2000},
  publisher={IEEE}
}

@article{oberg2007linear,
  title={Linear mixed effects models},
  author={Oberg, Ann L and Mahoney, Douglas W},
  journal={Topics in biostatistics},
  pages={213--234},
  year={2007},
  publisher={Springer}
}

@article{grudin2002computer,
  title={Computer-supported cooperative work: History and focus},
  author={Grudin, Jonathan},
  journal={Computer},
  volume={27},
  number={5},
  pages={19--26},
  year={2002},
  publisher={IEEE}
}

@inproceedings{begel2008pair,
  title={Pair programming: what's in it for me?},
  author={Begel, Andrew and Nagappan, Nachiappan},
  booktitle={Proceedings of the Second ACM-IEEE international symposium on Empirical software engineering and measurement},
  pages={120--128},
  year={2008}
}

@inproceedings{bigman2021pearprogram,
  title={PearProgram: A more fruitful approach to pair programming},
  author={Bigman, Maxwell and Roy, Ethan and Garcia, Jorge and Suzara, Miroslav and Wang, Kaili and Piech, Chris},
  booktitle={Proceedings of the 52nd ACM Technical Symposium on Computer Science Education},
  pages={900--906},
  year={2021}
}

@inproceedings{schenk2014distributed,
  title={Distributed-pair programming can work well and is not just distributed pair-programming},
  author={Schenk, Julia and Prechelt, Lutz and Salinger, Stephan},
  booktitle={Companion Proceedings of the 36th International Conference on Software Engineering},
  pages={74--83},
  year={2014}
}

@article{sapkota2025ai,
  title={Ai agents vs. agentic ai: A conceptual taxonomy, applications and challenges},
  author={Sapkota, Ranjan and Roumeliotis, Konstantinos I and Karkee, Manoj},
  journal={arXiv preprint arXiv:2505.10468},
  year={2025}
}

@inproceedings{villamor2018friends,
  title={Do friends collaborate and perform better?: A pair program tracing and debugging eye-tracking experiment},
  author={Villamor, Maureen and Rodrigo, MM},
  booktitle={18th Philippine Computing Science Congress},
  pages={9--16},
  year={2018}
}

@article{zhong2016impact,
  title={The impact of social factors on pair programming in a primary school},
  author={Zhong, Baichang and Wang, Qiyun and Chen, Jie},
  journal={Computers in Human Behavior},
  volume={64},
  pages={423--431},
  year={2016},
  publisher={Elsevier}
}

@book{bolboacua2021practical,
  title={Practical Remote Pair Programming: Best practices, tips, and techniques for collaborating productively with distributed development teams},
  author={Bolboac{\u{a}}, Adrian},
  year={2021},
  publisher={Packt Publishing Ltd}
}

@inproceedings{adeliyi2021investigating,
  title={Investigating remote pair programming in part-time distance education},
  author={Adeliyi, Adeola and Wermelinger, Michel and Kear, Karen and Rosewell, Jon},
  booktitle={Proceedings of the 2021 Conference on United Kingdom \& Ireland Computing Education Research},
  pages={1--7},
  year={2021}
}

@article{salleh2010empirical,
  title={Empirical studies of pair programming for CS/SE teaching in higher education: A systematic literature review},
  author={Salleh, Norsaremah and Mendes, Emilia and Grundy, John},
  journal={IEEE Transactions on Software Engineering},
  volume={37},
  number={4},
  pages={509--525},
  year={2010},
  publisher={IEEE}
}

@inproceedings{liu2024teaching,
  title={Teaching CS50 with AI: leveraging generative artificial intelligence in computer science education},
  author={Liu, Rongxin and Zenke, Carter and Liu, Charlie and Holmes, Andrew and Thornton, Patrick and Malan, David J},
  booktitle={Proceedings of the 55th ACM technical symposium on computer science education V. 1},
  pages={750--756},
  year={2024}
}

@article{kumar2025sharp,
  title={Sharp Tools: How Developers Wield Agentic AI in Real Software Engineering Tasks},
  author={Kumar, Aayush and Bajpai, Yasharth and Gulwani, Sumit and Soares, Gustavo and Murphy-Hill, Emerson},
  journal={arXiv e-prints},
  pages={arXiv--2506},
  year={2025}
}

@inproceedings{weisz2025examining,
  title={Examining the use and impact of an ai code assistant on developer productivity and experience in the enterprise},
  author={Weisz, Justin D and Kumar, Shraddha Vijay and Muller, Michael and Browne, Karen-Ellen and Goldberg, Arielle and Heintze, Katrin Ellice and Bajpai, Shagun},
  booktitle={Proceedings of the Extended Abstracts of the CHI Conference on Human Factors in Computing Systems},
  pages={1--13},
  year={2025}
}

@article{nosek1998case,
  title={The case for collaborative programming},
  author={Nosek, John T},
  journal={Communications of the ACM},
  volume={41},
  number={3},
  pages={105--108},
  year={1998},
  publisher={ACM New York, NY, USA}
}

@article{zhao2025generative,
  title={A Generative Artificial Intelligence (AI)-Based Human-Computer Collaborative Programming Learning Method to Improve Computational Thinking, Learning Attitudes, and Learning Achievement},
  author={Zhao, Gang and Yang, Lijun and Hu, Biling and Wang, Jing},
  journal={Journal of Educational Computing Research},
  pages={07356331251336154},
  year={2025},
  publisher={SAGE Publications Sage CA: Los Angeles, CA}
}

@article{wang2025impact,
  title={Impact of AI-agent-supported collaborative learning on the learning outcomes of University programming courses},
  author={Wang, Haoming and Wang, Chengliang and Chen, Zhan and Liu, Fa and Bao, Chunjia and Xu, Xianlong},
  journal={Education and Information Technologies},
  pages={1--33},
  year={2025},
  publisher={Springer}
}

@article{tang2024vizgroup,
  title={VizGroup: An AI-Assisted Event-Driven System for Real-Time Collaborative Programming Learning Analytics},
  author={Tang, Xiaohang and Wong, Sam and Pu, Kevin and Chen, Xi and Yang, Yalong and Chen, Yan},
  journal={arXiv preprint arXiv:2404.08743},
  year={2024}
}

@inproceedings{mirzakhmedova2024large,
  title={Are Large Language Models Reliable Argument Quality Annotators?},
  author={Mirzakhmedova, Nailia and Gohsen, Marcel and Chang, Chia Hao and Stein, Benno},
  booktitle={Conference on Advances in Robust Argumentation Machines},
  pages={129--146},
  year={2024},
  organization={Springer}
}

@inproceedings{ding2025multi,
  title={A Multi-Level Benchmark for Causal Language Understanding in Social Media Discourse},
  author={Ding, Xiaohan and Ping, Kaike and {\c{C}}ar{\i}k, Buse and Rho, Eugenia},
  booktitle={Proceedings of the 2025 Conference on Empirical Methods in Natural Language Processing},
  pages={28764--28778},
  year={2025}
}

@inproceedings{chen2024large,
  title={Is a large language model a good annotator for event extraction?},
  author={Chen, Ruirui and Qin, Chengwei and Jiang, Weifeng and Choi, Dongkyu},
  booktitle={Proceedings of the AAAI conference on artificial intelligence},
  volume={38},
  number={16},
  pages={17772--17780},
  year={2024}
}

@article{elo2008qualitative,
  title={The qualitative content analysis process},
  author={Elo, Satu and Kyng{\"a}s, Helvi},
  journal={Journal of advanced nursing},
  volume={62},
  number={1},
  pages={107--115},
  year={2008},
  publisher={Wiley Online Library}
}

@inproceedings{samei2014context,
  title={Context-based speech act classification in intelligent tutoring systems},
  author={Samei, Borhan and Li, Haiying and Keshtkar, Fazel and Rus, Vasile and Graesser, Arthur C},
  booktitle={International conference on intelligent tutoring systems},
  pages={236--241},
  year={2014},
  organization={Springer}
}

@article{nokes2015better,
  title={When is it better to learn together? Insights from research on collaborative learning},
  author={Nokes-Malach, Timothy J and Richey, J Elizabeth and Gadgil, Soniya},
  journal={Educational Psychology Review},
  volume={27},
  number={4},
  pages={645--656},
  year={2015},
  publisher={Springer}
}

@article{braun2023toward,
  title={Toward good practice in thematic analysis: Avoiding common problems and be (com) ing a knowing researcher},
  author={Braun, Virginia and Clarke, Victoria},
  journal={International journal of transgender health},
  volume={24},
  number={1},
  pages={1--6},
  year={2023},
  publisher={Taylor \& Francis}
}

@article{dixon2005synthesising,
  title={Synthesising qualitative and quantitative evidence: a review of possible methods},
  author={Dixon-Woods, Mary and Agarwal, Shona and Jones, David and Young, Bridget and Sutton, Alex},
  journal={Journal of health services research \& policy},
  volume={10},
  number={1},
  pages={45--53},
  year={2005},
  publisher={Sage Publications Sage UK: London, England}
}

@article{searle1980population,
  title={Population marginal means in the linear model: an alternative to least squares means},
  author={Searle, Shayle R and Speed, F Michael and Milliken, George A},
  journal={The American Statistician},
  volume={34},
  number={4},
  pages={216--221},
  year={1980},
  publisher={Taylor \& Francis}
}

@article{hagemann2023human,
  title={Human-AI teams—Challenges for a team-centered AI at work},
  author={Hagemann, Vera and Rieth, Mich{\`e}le and Suresh, Amrita and Kirchner, Frank},
  journal={Frontiers in Artificial Intelligence},
  volume={6},
  pages={1252897},
  year={2023},
  publisher={Frontiers Media SA}
}

@article{nitsch2024human,
  title={Human-centered approaches to AI-assisted work: the future of work?},
  author={Nitsch, Verena and Rick, Vera and Kluge, Annette and Wilkens, Uta},
  journal={Zeitschrift f{\"u}r Arbeitswissenschaft},
  volume={78},
  number={3},
  pages={261--267},
  year={2024},
  publisher={Springer}
}

@article{berretta2023defining,
  title={Defining human-AI teaming the human-centered way: a scoping review and network analysis},
  author={Berretta, Sophie and Tausch, Alina and Ontrup, Greta and Gilles, Bj{\"o}rn and Peifer, Corinna and Kluge, Annette},
  journal={Frontiers in Artificial Intelligence},
  volume={6},
  pages={1250725},
  year={2023},
  publisher={Frontiers Media SA}
}

@article{christensen2018cumulative,
  title={Cumulative link models for ordinal regression with the R package ordinal},
  author={Christensen, Rune Haubo B},
  journal={Submitted in J. Stat. Software},
  volume={35},
  pages={1--46},
  year={2018}
}

\appendix
\section{Appendix A: Large Language Model Prompt}
\vspace{0.5cm}
\begin{tcolorbox}[title=Prompt for Proactive Intervention and Direct Request, colback=white, colframe=black]
\small
You are Bob, an AI pair programming assistant focused on LEARNING.

\textcolor{black}{\textbf{Problem}}: \{problem\_title or 'General coding'\} - \{problem\_description\}

\textcolor{black}{\textbf{Language}}: \{programming\_language\}

\textcolor{black}{\textbf{Current code}}: \{code\_context or "No code visible yet"\}

\{ai\_history\_context\}

INTERVENTION APPROACH:
\begin{itemize}
\item Help users when they need it, but avoid unnecessary responses when they're satisfied
\item When users say 'I'm not sure', 'I need help', or ask questions, provide helpful guidance
\item When they say 'okay', 'thanks', 'got it', return "NO RESPONSE"
\item CRITICAL: Look at your recent messages above - you CANNOT repeat the same type of response
\item Each response must be MORE CONCRETE than your previous ones if they still need help
\item NEVER end responses with questions like "Need help with...?" or "Want me to...?"
\item Focus on brief conceptual hints rather than code snippets
\end{itemize}

Return EXACTLY "NO RESPONSE" (if no response needed) OR provide a helpful response (10-30 words).

\{current\_conversation\}
\end{tcolorbox}








\begin{tcolorbox}[title=Prompt for Task Scaffolding, colback=white, colframe=black]
\small
You are a coding tutor that creates minimal scaffolding to help students learn by doing.

A user wrote this comment in a \{language\} file:

"\{comment\_line.strip()\}"

Full code context:

\{full\_code\}

\textbf{CRITICAL RULES}:
\begin{enumerate}
\item Only provide scaffolding if the comment indicates the user wants to implement something
\item Generate MINIMAL scaffolding - just structure, NO solutions
\item Use descriptive TODO comments instead of \_\_ placeholders
\item Keep it SHORT (max 5-8 lines)
\item Students must fill in ALL the actual implementation
\end{enumerate}

\textbf{SCAFFOLDING REQUIREMENTS}:
\begin{itemize}
\item TODO comments
\item Clear, descriptive TODO guidance
\item NO actual implementation or solutions
\end{itemize}

\textbf{OUTPUT FORMAT}:
\begin{itemize}
\item If scaffolding needed: Return ONLY the minimal scaffolding code (NO markdown formatting, NO code blocks, just raw code)
\item If no scaffolding needed: Return exactly "NO SCAFFOLDING"
\item Do NOT use ```language``` formatting in your response
\item Return plain text code only
\end{itemize}

GOOD scaffolding examples:

Comment: "\# Create a function to calculate average"

\textbf{Output}: 
\begin{verbatim}
  # TODO: calculate sum of all numbers
  # TODO: divide sum by count of numbers
  # TODO: return the average
\end{verbatim}

Comment: "// Implement bubble sort"

Output:
\begin{verbatim}
  # TODO: loop through array multiple times
  # TODO: compare adjacent elements
  # TODO: swap if in wrong order
  # TODO: return sorted array
\end{verbatim}

BAD examples (too much solution):
\begin{itemize}
\item Any actual calculations or logic
\item Complete implementations
\item Specific values or algorithms
\end{itemize}
\end{tcolorbox}

\begin{tcolorbox}[title=Prompt for Code Run Feedback, colback=white, colframe=black]
\small
Code execution analysis:

Code: \{code\}

Problem: \{problem\_context or 'General coding'\}

Success: \{result.get('success', True)\}

Output: \{output if output else 'None'\}

Error: \{error if error else 'None'\}

CRITICAL RULES:
\begin{itemize}
\item Single loops through function results = EFFICIENT O(n)
\item ONLY suggest hashmap for actual nested for loops (for i, for j pattern)
\item Max-finding with single loop = CORRECT and EFFICIENT
\item ONLY analyze subtasks that have actual code implementation
\item Skip subtasks that are just comments, TODOs, or placeholders without code
\item Ignore commented-out subtask implementations
\item Ignore subtasks that only contain comments followed by "return None" or similar placeholders
\end{itemize}

Response (max 150 chars):
\begin{itemize}
\item Errors: "Fix: [issue]"
\item Wrong output: "Output: [issue]"
\item Actual inefficiency: "Optimize: [suggestion]"
\item Working efficiently: "correct"
\end{itemize}

Examples: "Fix: Missing )", "correct", 

"Subtask 1: correct, subtask 2: replace nested loops with hashmap"

(only mention subtasks with actual code)
\end{tcolorbox}

\begin{tcolorbox}[title=Prompt for Code Block Analysis, colback=white, colframe=black]
\small
Analyze this \{language\} code:

\{code\}

Problem: \{problem\_context.get('title', 'Unknown') if problem\_context else 'General coding'\}

CRITICAL RULES:
\begin{itemize}
\item Max-finding algorithms using single loops are EFFICIENT and CORRECT
\item ONLY suggest hashmap for actual nested loops (for i in range, for j in range pattern)
\item Single loop iterating through function results = O(n) = EFFICIENT
\item Don't flag: style, comments, missing subtasks, working algorithms
\end{itemize}

Only flag: undefined variables, syntax errors, actual logic bugs

JSON: \{"issue": \{"title": "...", "description": "...", "hint": "..."\}\}

Good code: \{"issue": \{"title": "Code looks good!", 
                      "description": "Correct and efficient.", 
                      "hint": "Well done!"\}\}
\end{tcolorbox}

\section{Appendix B: Programming Tasks during User Study}

\subsection{Task 1: Gift Card Purchase Assistant}
\textbf{Scenario:} A user has a gift card with a fixed value (e.g., \$100) and wants to buy two items from their shopping cart whose prices add up exactly to the gift card value.

\textbf{Subtasks:}
\begin{enumerate}
    \item Given a chosen item index, find another item whose price equals the remaining gift card balance.
    \item Return all pairs of item indices whose prices add up exactly to the gift card value \textit{(bonus: optimize to $O(n)$)}.
    \item Find the pair that includes the highest possible individual item price among all valid pairs.
\end{enumerate}




\subsection{Task 2: Step Tracker Insight}
\textbf{Scenario:} A fitness tracking app that tracks user daily step counts in an array.

\textbf{Subtasks:}
\begin{enumerate}
    \item Daily Average: Calculate the overall average steps per day across all recorded days.
    \item Best $k$-Day Streak Average: Find the highest average step count within any $k$ consecutive days \textit{(bonus: optimize to $O(n)$)}.
    \item Shortest Target Subarray: Find the shortest contiguous subarray where the total steps is at least a given target.
\end{enumerate}




\subsection{Task 3: Meeting Room Scheduler}
\textbf{Scenario:} A meeting room booking system that manages time slots for conference rooms. Each booking is represented as an interval [start\_time, end\_time].

\textbf{Subtasks:}
\begin{enumerate}
    \item Longest Meeting: Find the meeting with the longest duration.
    \item Detect Conflicts: Identify all pairs of meetings that have overlapping time intervals \textit{(bonus: optimize to $O(n \log n)$)}.
    \item Merge Overlapping: Merge all overlapping or adjacent meeting intervals into a single interval.
\end{enumerate}




\section{Appendix C: Questionnaire}

\subsection{Collaborative Learning Scale}
\textit{Note: Group here can either be other human + AI or AI partner.}

\begin{enumerate}
    \item I felt part of a learning community in my group.
    \item I actively exchanged my ideas with group members.
    \item I was able to develop new skills and knowledge from other members in my group.
    \item I was able to develop problem-solving skills through peer collaboration.
    \item Collaborative learning in my group was effective.
    \item Collaborative learning in my group was time-consuming.
    \item Overall, I am satisfied with my learning experience.
\end{enumerate}

\subsection{Social Presence Questionnaire}
\begin{enumerate}
    \item When I had real-time interaction during this programming task, I had my collaborator(s) in my mind’s eye (able to imagine how they were engaging in the task).
    \item During real-time interaction in this programming task, I felt that I was working with real partners, not with abstract or anonymous entities.
    \item Real-time interactions in this programming task could hardly be distinguished from face-to-face collaboration.
\end{enumerate}

\begin{table*}[h]
  \centering
  \caption{\revised{Frequency Distribution of Utterance Types Across Conditions and Post-Hoc Chi-Square Test Results}}
  \label{tab:utterance_frequencies}
  \begin{threeparttable}
    \begin{tabular}{
        l 
        r@{ }l 
        r@{ }l 
        r@{ }l 
        c c c
    }
      \toprule
      \multirow{2}[2]{*}{Utterance Type} & \multicolumn{6}{c}{Condition ($n$ / \%)} & \multicolumn{3}{c}{Statistical Test} \\
      \cmidrule(lr){2-7} \cmidrule(lr){8-10}
       & \multicolumn{2}{c}{HHAI-S} & \multicolumn{2}{c}{HHAI-P} & \multicolumn{2}{c}{HAI} & $\chi^2(df=2)$ & $V$ & $p_{adj}$ \\
      \midrule
      \textbf{Question (Seek Info)} & \textbf{174} & \textbf{(12.06)} & \textbf{190} & \textbf{(14.22)} & \textbf{126} & \textbf{(35.10)} & \textbf{119.24} & \textbf{.19} & \textbf{$< .001^{***}$} \\
      
      Question (Seek Confirm)       & 80           & (5.54)           & 76           & (5.69)           & 20           & (5.57)           & 0.03            & .00          & 1.000           \\
      
      Answer / Reply                & 27           & (1.87)           & 31           & (2.32)           & 1            & (0.28)           & 6.40            & .05          & .205            \\
      
      Strategy Proposal             & 60           & (4.16)           & 71           & (5.31)           & 14           & (3.90)           & 2.58            & .03          & 1.000           \\
      
      Implementation Proposal       & 253          & (17.53)          & 201          & (15.04)          & 63           & (17.55)          & 3.46            & .03          & 1.000           \\
      
      \textbf{Elaboration / Justification} & \textbf{77} & \textbf{(5.34)} & \textbf{35} & \textbf{(2.62)} & \textbf{0} & \textbf{(0.00)} & \textbf{29.88} & \textbf{.10} & \textbf{$< .001^{***}$} \\
      
      Think-Aloud (Cognitive)       & 287          & (19.89)          & 311          & (23.28)          & 65           & (18.11)          & 7.00            & .05          & .251            \\
      
      Read-Aloud / Verbalization    & 22           & (1.53)           & 12           & (0.90)           & 4            & (1.11)           & 2.31            & .03          & 1.000           \\
      
      \textbf{Acknowledgment/Acceptance} & \textbf{237} & \textbf{(16.42)} & \textbf{193} & \textbf{(14.45)} & \textbf{24} & \textbf{(6.69)} & \textbf{22.03} & \textbf{.08} & \textbf{$< .001^{***}$} \\
      
      Disagreement / Rejection      & 41           & (2.84)           & 37           & (2.77)           & 4            & (1.11)           & 3.59            & .03          & 1.000           \\
      
      Coordination / Turn-Taking    & 76           & (5.27)           & 59           & (4.42)           & 11           & (3.06)           & 3.44            & .03          & 1.000           \\
      
      Affective Expression          & 109          & (7.55)           & 120          & (8.98)           & 27           & (7.52)           & 2.11            & .03          & 1.000           \\
      \bottomrule
    \end{tabular}
    \begin{tablenotes}[para, flushleft]
      \small
      \textit{Note.} Values in parentheses indicate column percentages. HHAI-P = HHAI-Personal; HHAI-S = HHAI-Shared. $V$ represents Cramer's V effect size. $p_{adj}$ represents p-values adjusted for multiple comparisons using the Holm-Bonferroni method. Bold rows indicate statistical significance ($^{***}p < .001$).
    \end{tablenotes}
  \end{threeparttable}
\end{table*}

\begin{table*}[h]
\centering
\caption{\revised{Accuracy of GPT-4o-mini annotations across conditions.}}
\begin{tabular}{lccc}
\hline
\textbf{Condition} & \textbf{User Studies} & \textbf{User Utterances} & \textbf{Accuracy (\%)} \\
\hline
HHAI-Shared   & 3 & 515 & 88.54 \\
HHAI-Personal & 3 & 461 & 90.89 \\
HAI           & 3 & 151 & 82.78 \\
\hline
\end{tabular}

\label{tab:annotation_accuracy}
\end{table*}

\subsection{Additional Questionnaire}
\begin{enumerate}
    \item I felt responsible for understanding the AI’s suggestions before applying them.
    \item The AI felt too interruptive/disruptive.
\end{enumerate}

\revised{\section{Appendix D: Supplementary Analysis on User Utterances}\label{Appendix:D}} 
\revised{To understand how participants engaged throughout the session and to further analyze their interactions, we annotated a randomized subset of user utterances (3,138 utterances across all studies). Our annotation process is as follows:
\begin{enumerate}
    \item To construct the annotation scheme, one author conducted initial open coding on $\sim\!10\%$ of the utterances to identify recurring patterns \cite{elo2008qualitative}. The initial codes were then reviewed and discussed with two other authors, and through this process, we finalized the annotation scheme presented in the table \ref{tab:combined_annotation_scheme}.
    \item We then annotated three sampled user studies (3 studies across 3 conditions = 1127 user utterances = $\sim\!36\%$ of the data). Two annotators independently coded this subset (Cohen’s $\kappa = 0.853$) and resolved all disagreements to produce final labels.
    \item To annotate the remaining portion of the dataset ($\sim\!64\%$), we used the gpt-4o-mini model to classify each utterance using our annotation scheme, building on prior work in AI-assisted annotation \cite{mirzakhmedova2024large, ding2025multi, chen2024large}. To provide additional context \cite{samei2014context}, the model was given the three preceding and three following utterances as the context window for each target instance. We evaluated the model by comparing its predictions on our human-annotated subset and reported its resulting accuracy (table. \ref{tab:annotation_accuracy}).
    \item Finally, we calculated and reported the proportion of each utterance type across all conditions, as presented in the table \ref{tab:combined_annotation_scheme}.
\end{enumerate}}

\end{document}